\documentclass[a4paper,fleqn,usenatbib]{mnras} 

\pdfoutput=1
\usepackage[T1]{fontenc}
\usepackage{ae,aecompl}
\usepackage[export]{adjustbox}
\usepackage{graphicx}
\usepackage{times}

\usepackage[pscoord]{eso-pic}
\usepackage{amsmath}
\usepackage{amssymb}

\newcommand{\msun}{{\,\rm M_\odot}}

\def\gsim{ \lower .75ex \hbox{$\sim$} \llap{\raise .27ex \hbox{$>$}} }
\def\lsim{ \lower .75ex \hbox{$\sim$} \llap{\raise .27ex \hbox{$<$}} }

 \title[LG star formation in WDM and SIDM]{Local Group star formation in warm and self-interacting dark matter cosmologies}
\author[M.~R.~Lovell et al.]{Mark R. Lovell\thanks{email: lovell@hi.is}$^{1,2}$, Wojciech Hellwing$^{3}$, Aaron Ludlow$^{4}$, Jes\'us Zavala$^{1}$, \newauthor Andrew Robertson$^{2}$, Azadeh Fattahi$^{2}$, Carlos S. Frenk$^{2}$ and Jennifer Hardwick$^{4}$ \\
$^{1}$Center for Astrophysics and Cosmology, Science Institute, University of Iceland, Dunhagi 5, 107 Reykjavik, Iceland \\
$^{2}$Institute for Computational Cosmology, Durham University, South Road, Durham DH1 3LE, UK \\
$^{3}$Center for Theoretical Physics, Polish Academy of Sciences, Al. Lotnik\'ow 32/46, 02-668 Warsaw, Poland\\
$^{4}$ICRAR M468, The University of Western Australia, 35 Stirling Hwy, Crawley, Western Australia, 6009}

\date{Accepted ... Received ...; in original form ...} 
\pubyear{2019}

\begin{document}

\label{firstpage}
\pagerange{\pageref{firstpage}--\pageref{lastpage}} 
  
\maketitle

\begin{abstract}

\noindent The nature of the dark matter can affect the collapse time of dark matter haloes, and can therefore be imprinted in observables such as the stellar population ages and star formation histories of dwarf galaxies. In this paper we use high resolution hydrodynamical simulations of Local Group-analogue (LG) volumes in cold dark matter (CDM), sterile neutrino warm dark matter (WDM) and self-interacting dark matter (SIDM) models with the EAGLE galaxy formation code to study how galaxy formation times change with dark matter model. We are able to identify the same haloes in different simulations, since they share the same initial density field phases.  We find that the stellar mass of galaxies depends systematically on resolution, and can differ by as much as a factor of two in haloes of a given dark matter mass. The evolution of the stellar populations in SIDM is largely identical to that of CDM, but in WDM early star formation is instead suppressed. The time at which LG haloes can begin to form stars through atomic cooling is delayed by $\sim$200~Myr in WDM models compared to CDM. It will be necessary to measure stellar ages of old populations to a precision of better than 100~Myr, and to address degeneracies with the redshift of reionization -- and potentially other baryonic processes --  in order to use these observables to distinguish between dark matter models.   

\end{abstract}


\section{Introduction}
\label{intro}

The nature of dark matter remains one of the biggest unanswered questions in physics. The simplest picture of dark matter, in which it constitutes a collisionless dust that interacts with baryons via gravity alone and has a negligible thermal velocity dispersion, has matched observations of the cosmic microwave background \citep[CMB, e.g.][]{wmap1,Planck16} and, together with dark energy, has successfully predicted the large scale structure of the low redshift Universe \citep{Cole05,Eisenstein05}. This model of dark energy plus cold dark matter (CDM) now serves as the basis for simulations of galaxy formation. These simulations model a rich set of astrophysics processes, including subgrid recipes for supernova feedback, active galactic nuclei, and magnetic fields; and now produce galaxy properties that resemble of observed systems in our own Universe \citep[e.g.][]{dubois14,vogelsberger14,Schaye15,Sawala16a,Pillepich17}. To the extent of current observations, the CDM model is a viable model of our Universe.  

The largest remaining obstacle to adopting the CDM hypothesis as an accurate theory is the failure to detect the CDM particle. This particle is typically assumed to be a supersymmetric, weakly interacting massive particle \citep[WIMPs;][]{Ellis_84}, and efforts to find evidence for this class of particles in collider searches \citep{Aaboud18,Sirunyan18}, direct detection experiments \citep{Akerib17,Aprile18} and indirect detection observations \citep{Gaskins16,Slatyer17} have all returned null results. Another proposed particle, the QCD axion, is also a CDM candidate \citep{Turner91}, but attempts to detect this particle have also failed to find sufficient evidence for its existence in the quantities needed to be the dark matter \citep{Du18,Braine19}.

Further alternatives to WIMPs can make predictions for structure formation that are significantly different to CDM, since they can be non-classical, undergo collisions and have a thermal velocity dispersion. These alternatives include axion-like particles that behave as fuzzy dark matter \citep[e.g.][]{Marsh16,Mocz17}, self-interacting dark matter (SIDM) particles \citep{Spergel00}, particles that interact with relativistic standard model particles \citep{Boehm14,Schewtschenko15} or dark photons \citep{Buckley14,Vogelsberger16}, and sterile neutrinos acting as warm dark matter (WDM) \citep{Dodelson94,Shi99,Laine08,Lovell16}, the latter of which may have already been detected in X-ray emission at an energy of 3.55~keV \citep{Boyarsky14a,Bulbul14,boyarsky14c,Ruchayskiy15,Cappelluti17,Boyarsky18,Hofmann19}, although this interpretation is disputed \citep{Anderson14,Jeltema14,Shah16}. Exploring these alternatives provides us with extra avenues of enquiry as to which particles are cosmologically viable, and may then be followed up with dedicated indirect searches of particle physics experiments.

The nature of dark matter can be imprinted on multiple observables, many of which can broadly be divided into three categories: gravitational lensing anomalies in group haloes, high-redshift gas physics, and the properties of local dwarf galaxies. Lensing anomalies are further broken down into anomalies in lensing arcs \citep{Vegetti14,Hezaveh16,Li16} and flux anomalies \citep{Mao98,Dalal02,Metcalf02,XuD09}: the former are expected to provide constraining power on the dark matter in future surveys \citep{Li16,Despali19a,Despali19b}, whereas current flux anomalies data have been claimed to have already placed strong constraints \citep{Hsueh19,Gilman20}. The second regime is of high-redshift gas physics, including both competitive constraints from the statistics of the Lyman-$\alpha$ forest (\citealp{Viel05,Viel13,Baur16,Irsic17,PalanqueDelabrouille19}; but see also \citealp{Garzilli18,Garzilli19}) and the timing of the reionization epoch \citep{Bose16c,Lovell18a}.

These two regimes are bridged, perhaps surprisingly, by the local dwarf galaxy observables category. First, local dwarfs share with lensing measurements information about the abundance of dark matter haloes with masses $<10^{10}\msun$ \citep{Bode01,Schneider2012,Bose16a}, plus their structure \citep{Spergel00,Lovell12,Bose16c,Bozek16,Ludlow16,Vogelsberger16}, which is also relevant for lensing studies \citep{Gilman19}; for a recent review on different dark matter models and their impact on structure formation see \citet{Zavala19b}. Recent attempts to measure the Milky Way (MW) halo substructure content with gaps in stellar streams  \citep{Ibata02,Johnston02} have reported strong constraints on WDM models \citep{Banik19}, although the translation between these limits -- plus the lensing constraints -- and particle physics models is unclear \citep{Lovell20}. Second, the high redshift gas constraints are related to when dark matter haloes first began to collapse and form stars, particularly with respect to the reionization constraints; structure formation occurs later in the presence of a power spectrum cut-off \citep{Lovell12,Schultz14,Bose16c} and therefore we can expect that the ages of the oldest stellar populations may bear the imprint of the dark matter particle free-streaming length. 


 This effect on the oldest stellar populations has been shown in hydrodynamical simulations. \citet{Governato15} showed that the initial onset of star formation in a single isolated dwarf galaxy was delayed by up to 2~Gyr in a 2~keV thermal relic-WDM cosmology compared to in CDM. Similar sets of exercises were taken by \citet{Bozek19} and \citet{Maccio19} with the benefit of larger samples of isolated haloes, and \citet{Lovell17b} found evidence for later star formation in lower resolution Local Group (LG) analogue environments; these three more recent studies also employed better constrained dark matter models than \citet{Governato15}.

 An alternative approach is to consider reionization-epoch galaxies, where the apparent delay constitutes a larger proportion of the cosmic age of galaxies relative to the LG population. \citet{Liu19} ran tailored high redshift galaxy models of a WDM 3~keV thermal relic, including a model of population III stars, and \citet{Lovell19b} showed  using a larger, lower resolution volume that galaxies in the ETHOS framework \citep{CyrRacine16,Vogelsberger16} -- which features a WDM-like power spectrum cut-off due to interactions between the dark matter and dark radiation -- exhibit similar behaviour: both studies found a delay of $\sim$200~Myr in the onset of WDM star formation. This delay is shorter than the 1-2~Gyr found by \citet{Governato15}, which is likely due to the longer free-streaming length associated with the 2~keV model than that applied in the two subsequent studies, although differences in the application of astrophysics cannot be ruled out. We infer that it may therefore be possible to discern the nature of dark matter from the stellar populations of both the highest redshift galaxies and possibly from the fossil galaxies from that era that remain in the LG. 

In this paper we introduce a set of high resolution resimulations of the APOSTLE simulations of LG-analogue volumes \citep{Fattahi16,Sawala16a}, exploring a range of alternative dark matter models, one SIDM and three WDM. This set of simulations is drawn from a common set of initial conditions, which enables us to compare the models on a halo-by-halo basis, even measuring the effect of changing the mass resolution, due to the the application of the halo matching method developed in \citet{Lovell14} and \citet{Lovell18b}. We also present higher resolution simulations than were available in \citet{Lovell17b} but still with the benefit of the LG setting, thus modelling satellites rather than isolated dwarfs as has been the case in other alternative dark matter comparisons \citep{robles2017,Fitts19}. We begin our analysis of these simulations with a discussion of the star formation histories, including the initial onset of star formation.

 This paper is organised as follows. In Section~\ref{sims} we present our simulations and methods, in Section~\ref{res} we show our results, and we draw our conclusions in Section~\ref{conc}.

\section{Simulations and methods} 
\label{sims}

The simulations used in this study are all either part of, or derived from, the APOSTLE project \citep{Fattahi16,Sawala16a}. This original suite comprised simulations of 12 Local-Group analogue systems (labelled AP-1 through AP-12) in the CDM cosmology, featuring analogues of the MW and M31 galaxy dark matter haloes. In this study we consider seven simulations from the original APOSTLE suite: medium resolution (MR) simulations of six APOSTLE volumes (AP-1--AP-6) plus one high resolution (HR) simulation of one volume, AP-1 \footnote{In \citet{Sawala16a} and \citet{Fattahi16} the resolution levels labelled here as HR and MR are instead labelled L1 and L2 respectively.}.

Our simulations were performed with the same code and galaxy formation model as the original APOSTLE runs \citep{Schaye15,Crain15}, namely a modified version of {\sc p-gadget3} \citep{Springel08b}. This model features cooling, star formation \citep{Schaye08}, black hole growth \citep{Springel05c,RosasGuevara15}, supernova feedback and AGN feedback \citep{Booth09,DallaVecchia12}. We use the REFERENCE (hereafter, `Ref') models parameters for the WDM runs, irrespective of the resolution, as was also the case for CDM APOSTLE. In this model, hydrogen reionization is modelled as a spatially uniform and time dependent background radiation field as described by \citet{Haardt01,Wiersma09}, and is switched on at $z=11.5$. We also use a version of the AP-1-MR run in which the redshift of reionization was set to $z_\rmn{re}=7$ (`Rec7') instead of the fiducial $z_\rmn{re}=11.5$ (Fattahi~et~al., in prep.) We note that the subgrid model employs a temperature floor of 8000~K, which will affect the smallest halo mass that can form a galaxy \citep{Tegmark97,Abel02,Smith15}; we will discuss further limitations of the subgrid models in Section~\ref{sec:oosf}.

To the original APOSTLE suite we add simulations run with three WDM models and one SIDM model. Each WDM model represents a sterile neutrino particle, which is described by two parameters: its mass, $M_\rmn{s}$, and a lepton asymmetry, $L_{6}$. These two parameters combine to set the sterile neutrino decay rate and also the shape of the linear matter power spectrum (see \citealp{Lovell16} for a complete description)\footnote{Sterile neutrinos form a subset of WDM models. In this study we refer to our sterile neutrino models as `WDM' for brevity.}. The three WDM models adopt lepton asymmetry values $L_{6}=9$, 10, and 11.2, and all three have $M_\rmn{s}=7$~keV. The $L_6=11.2$ model (hereafter labelled as LA11) is the model with the lowest-$k$ cut-off permitted by the 3.55~keV line and is therefore the most distinct from CDM; $L_6=9$ (LA9) has the highest $k$ cut-off in agreement with the line, and $L_6=10$ (LA10) is an intermediate case.

In this study we analyse the six MR LA10 volumes (AP-1--AP-6) from \citet{Lovell17b}, and the LA11 AP-1 runs from \citet{Lovell19a} (both HR and MR), plus three previously unpublished simulations: an LA9 AP-1 volume (HR only) and also an SIDM model run on the same volume with a velocity independent transfer cross-section per unit mass of $10~\rmn{cm^{2}g^{-1}}$, which we label SIDM10 (HR and MR). The implementation of particle-on-particle scattering in the EAGLE code is described  in \citet{Robertson18}, based on the method of \citet{RobertsonA17}. The SIDM10 simulation differs from its CDM counterpart in that it was run with the RECAL (hereafter, `Rec') calibration parameters, which is optimized for simulations with gas mass resolution $\sim10^{5}\msun$. Unlike either the standard Rec or Ref models, SIDM10 used a \citet{Haardt01} reionization field that switched on at $z=9$ rather than $z=11.5$; we refer to this composite model as `Rec9'.  The properties of all our simulations are summarized in Table~\ref{tab:sims}. Note that, although each simulation has a long form name given in the table, because we are typically using versions of one volume and one resolution (HR), for brevity we will refer to each simulation just by its dark matter model, e.g. we shorten `AP-HR-CDM-Ref' to `CDM' and `AP-HR-SIDM10-Rec9' to `SIDM10', except in figure legends where we instead use CDM-Ref and SIDM10-Rec9, respectively.

Haloes are identified using the friends-of-friends (FoF) algorithm, with a linking length $b=0.2$. Subhaloes are then extracted from the FoF haloes using the {\sc subfind} gravitational unbinding code \citep{Springel01a,Dolag09}.  The minimum number of particles required for a subhalo to be included in the catalogue is 20. The properties of subhaloes located close to the edge of the high-resolution zoom region will be altered unphysically due to the local presence of heavy, low-resolution particles whose purpose is to model the large-scale tidal fields accurately. Therefore, we only analyse haloes located less than 3~Mpc (physical) from the barycentre of the M31-MW system except where stated otherwise.

The fact that both the dark matter model and the galaxy formation parameters are changed simultaneously between the CDM and SIDM10 simula2tions complicates the interpretation of the results: we have therefore run a version of the CDM volume with the same galaxy formation model and reionization parameters as the SIDM10 simulations. We have run this new CDM volume down to redshift $z=5$; we will mostly focus on the properties of galaxies formed in the first gigayear which will be less influenced by the cumulative effect of the different feedback parameters between Ref and Rec.

Where two simulations are performed of the same initial phases, e.g.  AP-HR-CDM and AP-MR-CDM, or  AP-HR-CDM and AP-HR-LA11, we use the Lagrangian matching method developed by \citet{Lovell14} and later expanded by \citet{Lovell18b} to match haloes between simulations. To summarize, for subhaloes we identify the dark matter particles that the subhalo had at its peak mass over the history of the simulation, (or has at $z=0$ for isolated haloes) and compute which halo in the counterpart simulation has the same Lagrangian distribution of particles in the initial conditions.  The quality of each match is expressed in a merit parameter, $R$, where $R=1$ denotes that the Lagrangian patches of the two haloes overlap exactly, and $R<0.5$ instead implies that the two haloes are not the same object, or at least have had very divergent formation histories in the two simulations. In this study we include matches for which $R>0.9$. The one exception to the infall particle choice is our generation of matched catalogues between AP-HR-CDM and AP-HR-CDM-Rec9, where we instead match all of the haloes present at $z=6$ using their $z=6$ dark matter particles.

Crucially, this method does not rely on the two counterpart simulations sharing the same particle ID schemes, or even the same number of particles, and is therefore suitable for resolution studies, such as those performed in this paper. Finally, this algorithm is also able to match subhaloes between snapshots in the same simulation, and can therefore be used to build robust merger trees down to very early times: we will use this method in Section~\ref{sec:oosf} to determine the epoch at which LG satellites are first able to undergo HI cooling.

\begin{figure*}
    \centering
    \setbox1=\hbox{\includegraphics[scale=0.23]{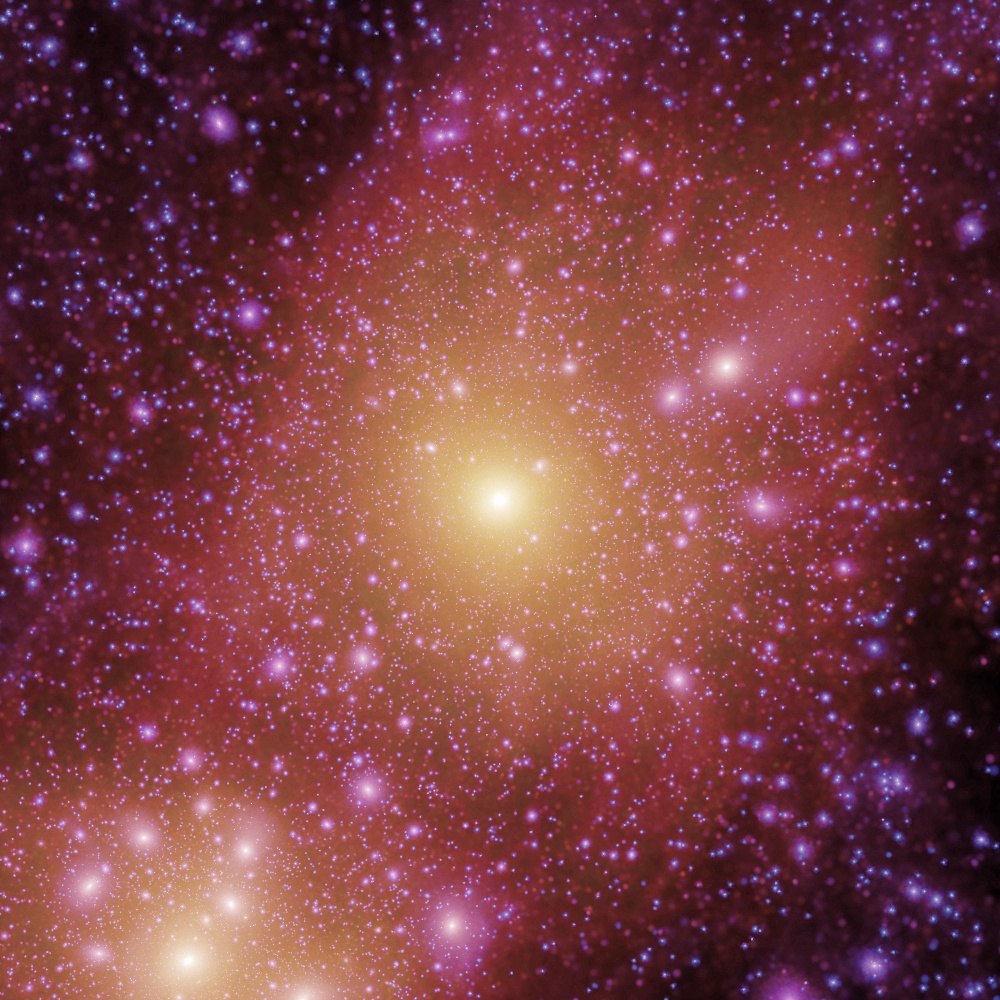}}
    
     \includegraphics[scale=0.23]{V1HRCDM_127M31_BSize0p6Mpc2_X1Y2_001.jpg}\llap{\makebox[\wd1][l]{\raisebox{0.8\wd1}{\includegraphics[width=0.28\textwidth]{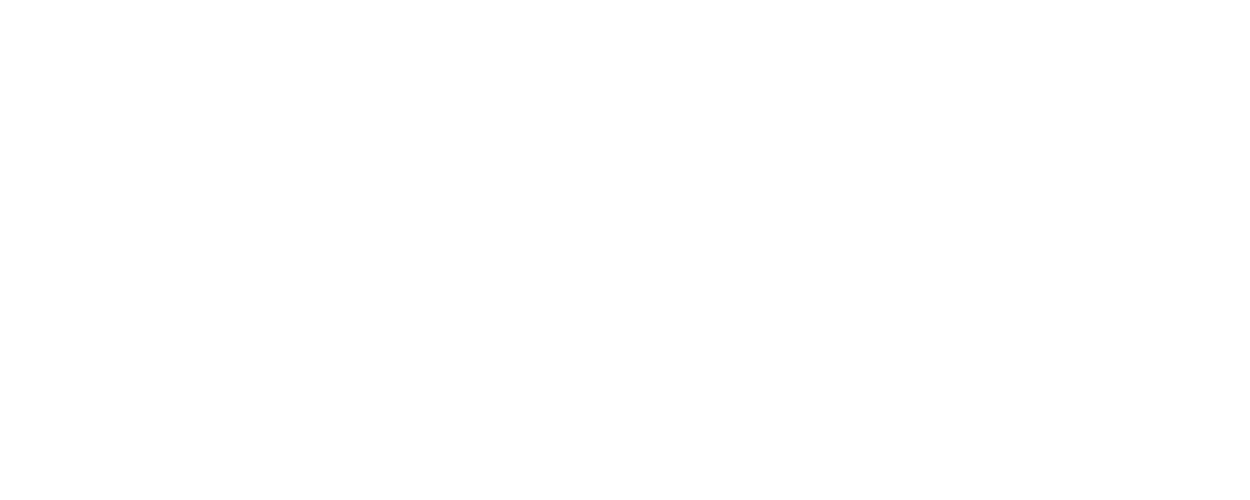}}}}
      \includegraphics[scale=0.23]{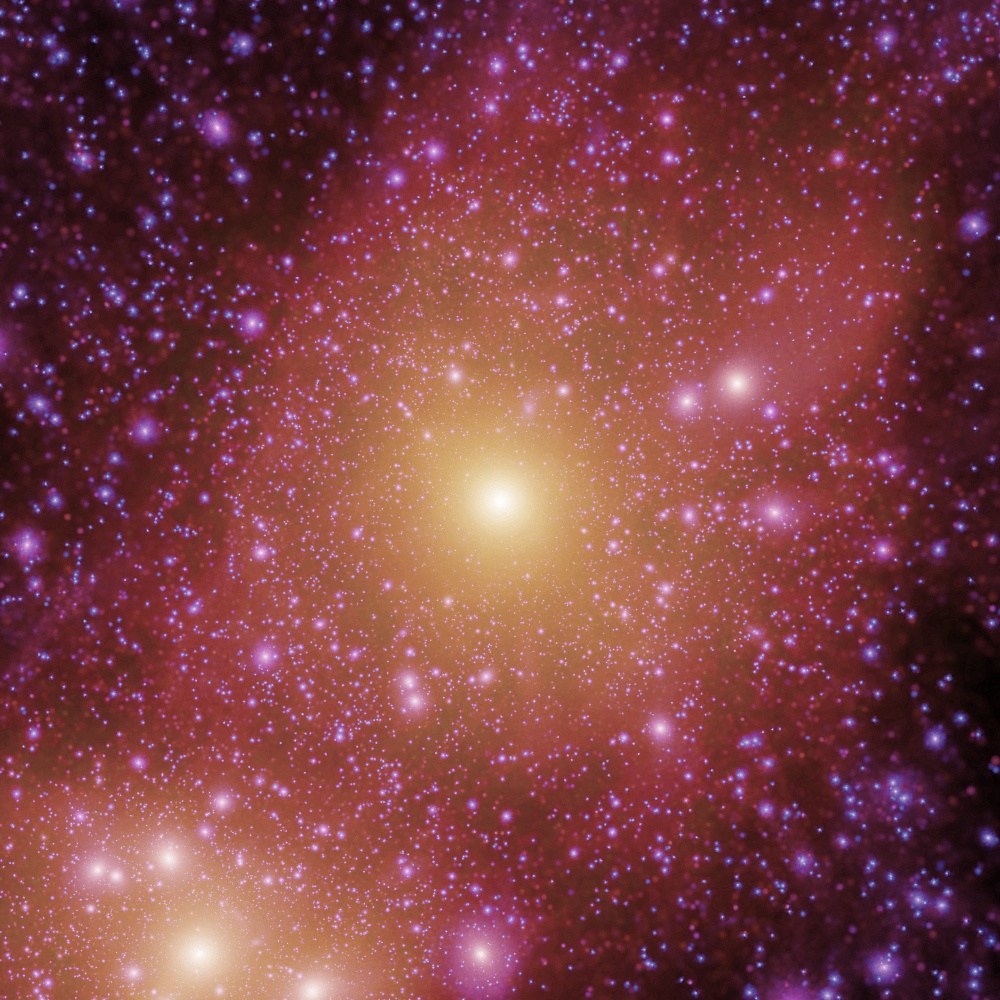}\llap{\makebox[\wd1][l]{\raisebox{0.8\wd1}{\includegraphics[width=0.28\textwidth]{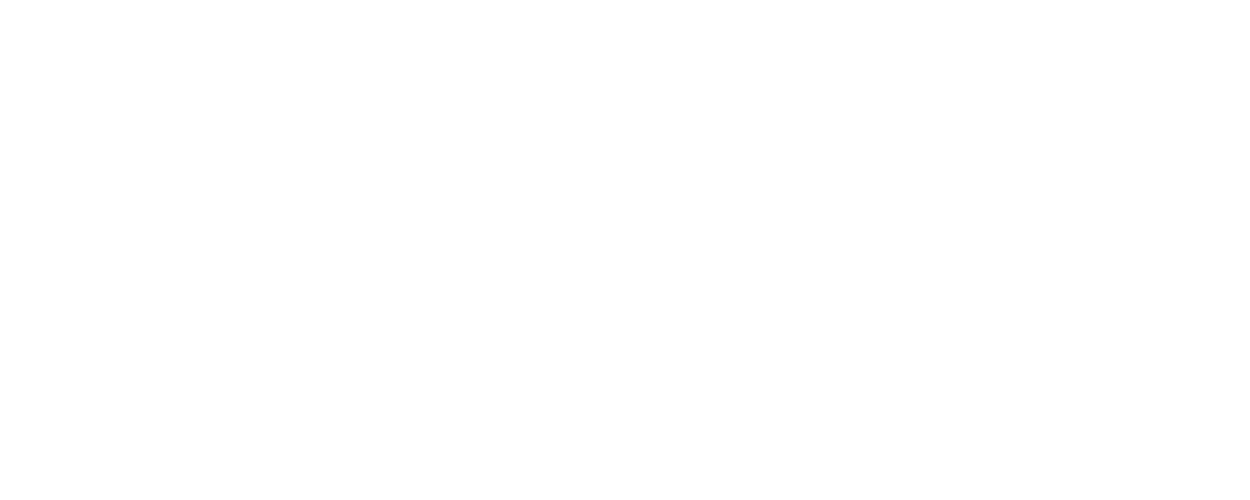}}}} \\
         \vspace{-3mm}
      \includegraphics[scale=0.23]{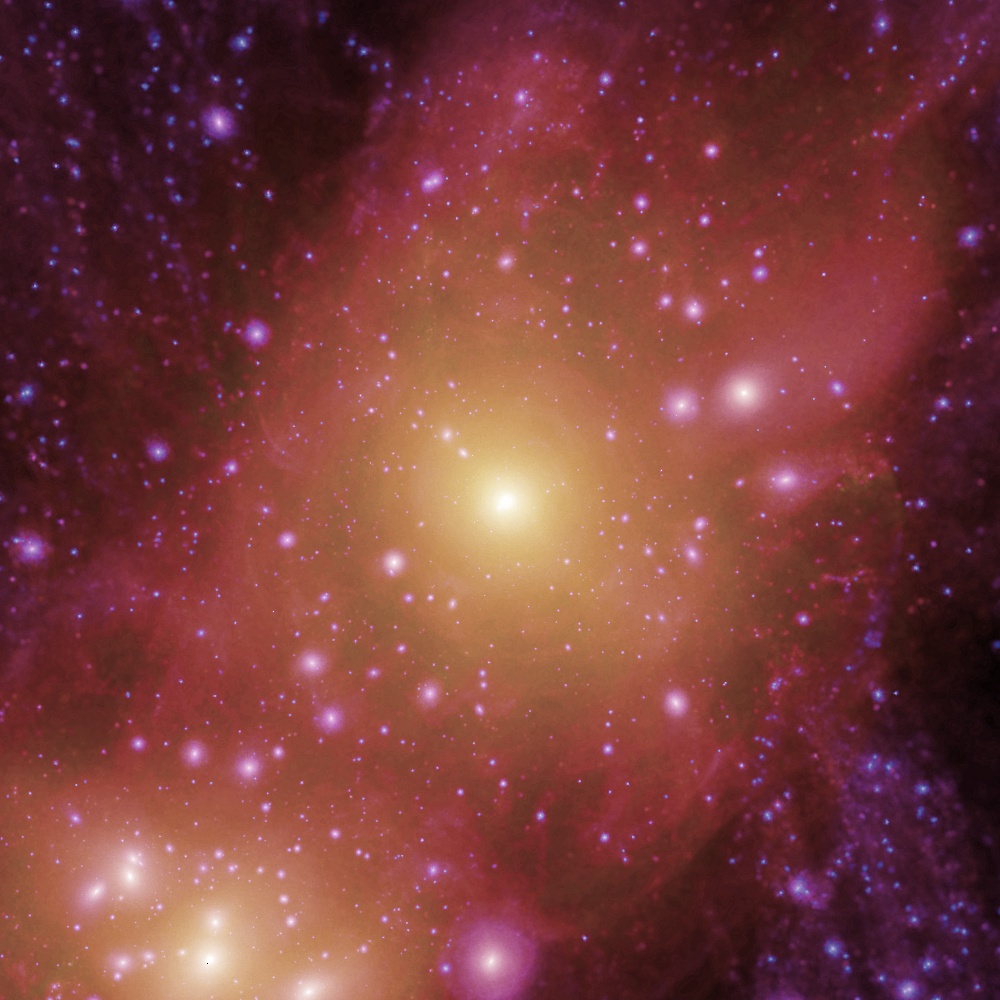}\llap{\makebox[\wd1][l]{\raisebox{0.8\wd1}{\includegraphics[width=0.28\textwidth]{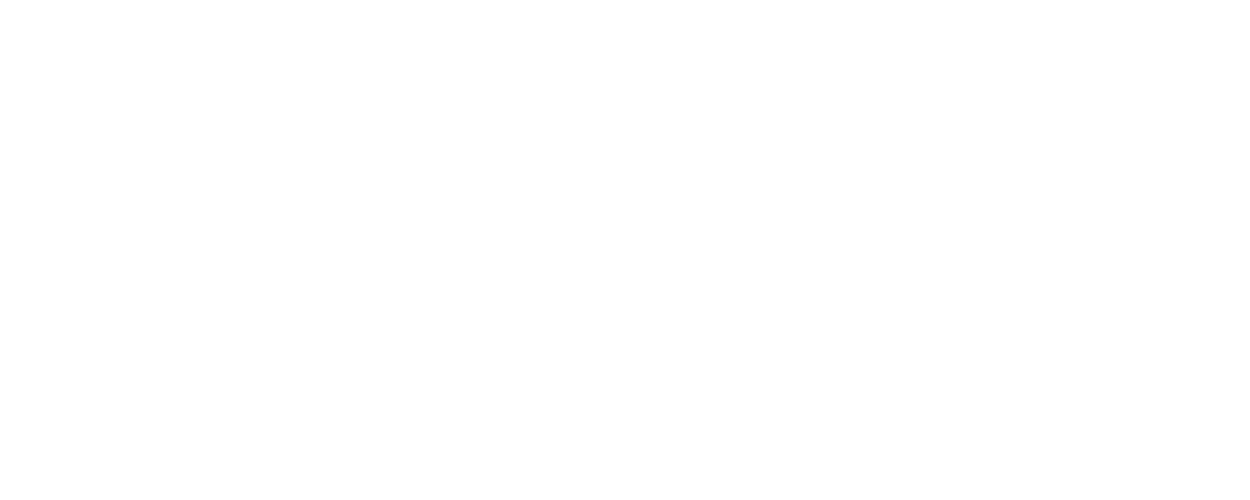}}}}
      \includegraphics[scale=0.23]{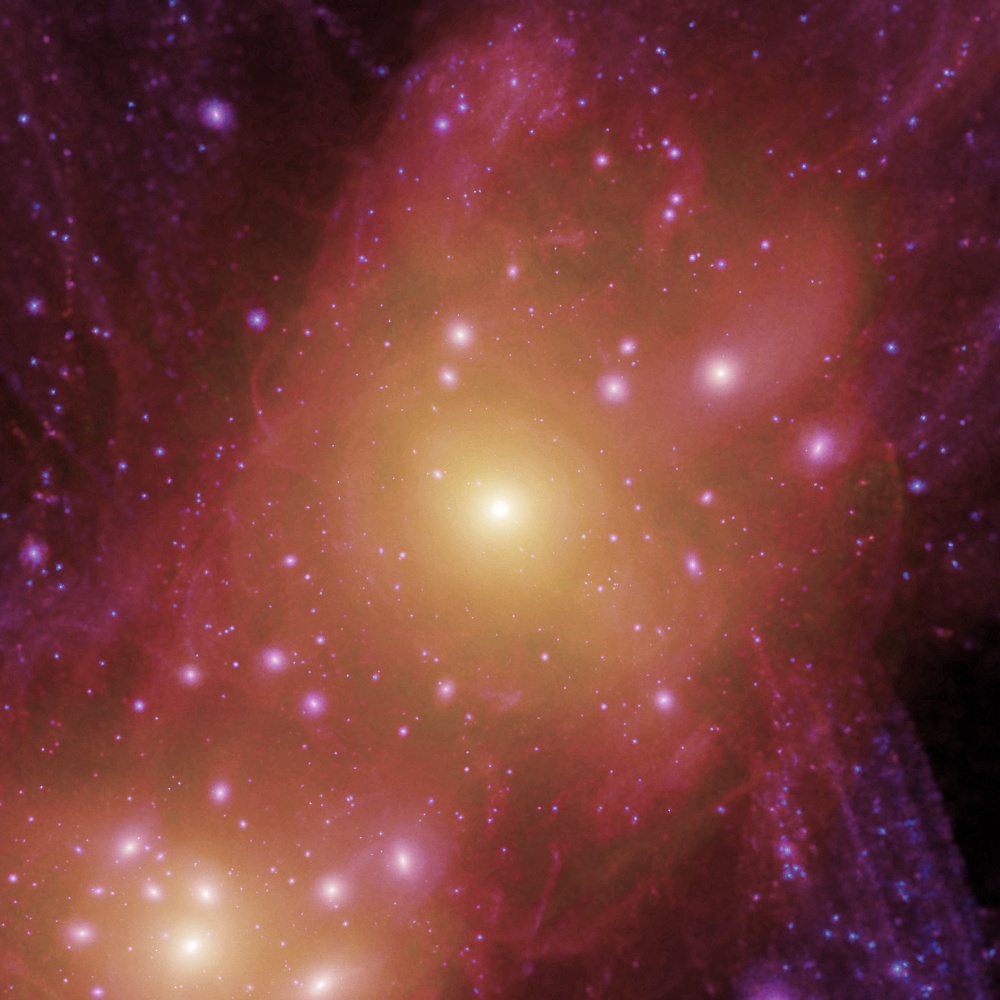}\llap{\makebox[\wd1][l]{\raisebox{0.8\wd1}{\includegraphics[width=0.28\textwidth]{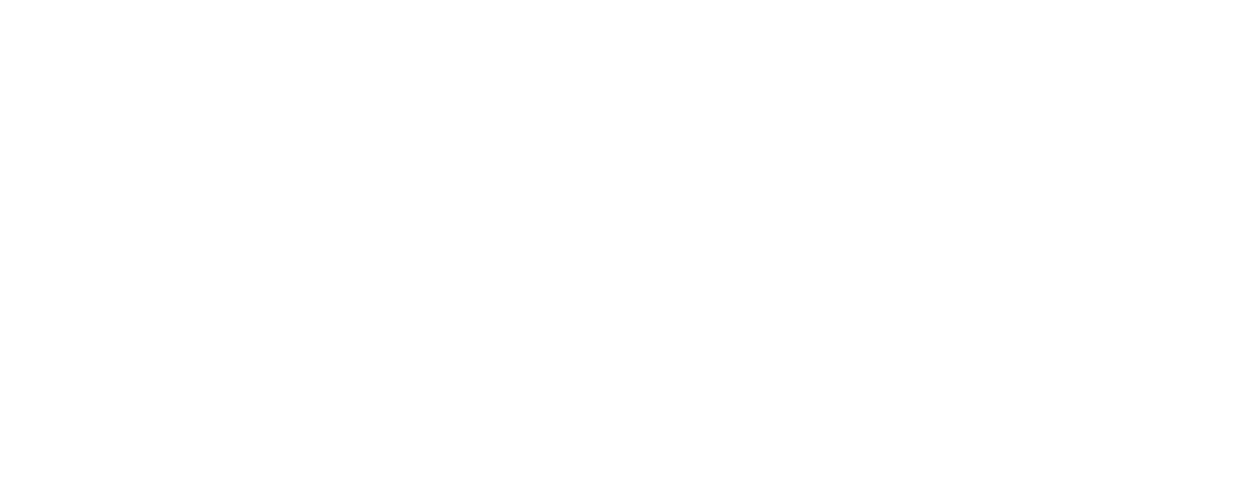}}}} \\

    \caption{The projected $z=0$ dark matter density in our four high resolution simulations, centred on the M31-analogue in each simulation. The four panels are AP-HR-CDM (top left), AP-HR-SIDM10-Rec9 (top right), AP-HR-LA9 (bottom left) and AP-HR-LA11 (bottom right). The image intensity indicates the projected density of dark matter, and the hue is related to the velocity dispersion. Each image is 1.2~Mpc on a side, and the density is projected over a distance of 0.8~Mpc -- slightly more than three times the halo radius -- both in front of and behind the image centre.}
    \label{fig:DM_Images}
\end{figure*}

\begin{figure*}
    \centering
    
     \setbox1=\hbox{\includegraphics[scale=0.23]{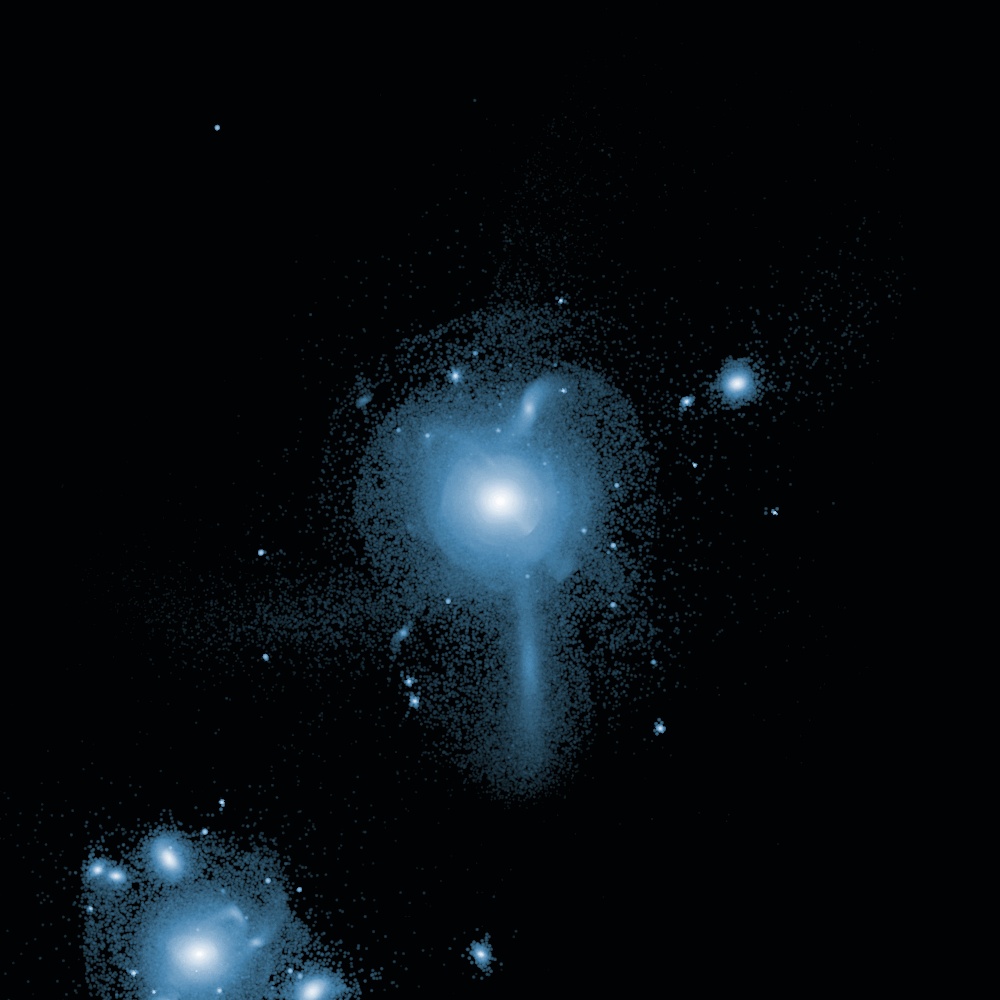}}
    
     \includegraphics[scale=0.23]{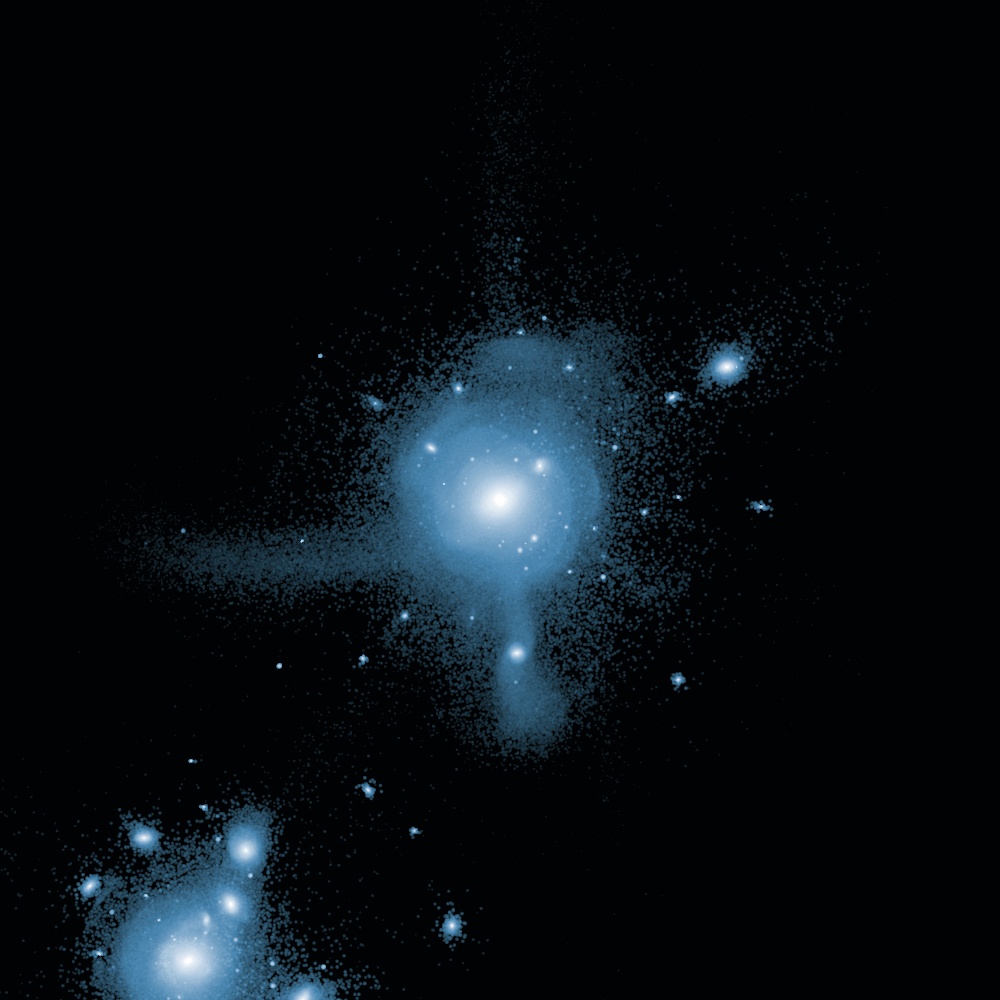}\llap{\makebox[\wd1][l]{\raisebox{0.8\wd1}{\includegraphics[width=0.28\textwidth]{CDM_Label-eps-converted-to.pdf}}}}
      \includegraphics[scale=0.23]{Stars_V1HRX10_127M31_BSize0p6Mpc2_X1Y2_001.jpg}\llap{\makebox[\wd1][l]{\raisebox{0.8\wd1}{\includegraphics[width=0.28\textwidth]{SIDM10_Label-eps-converted-to.pdf}}}} \\
        \vspace{-3mm}
      \includegraphics[scale=0.23]{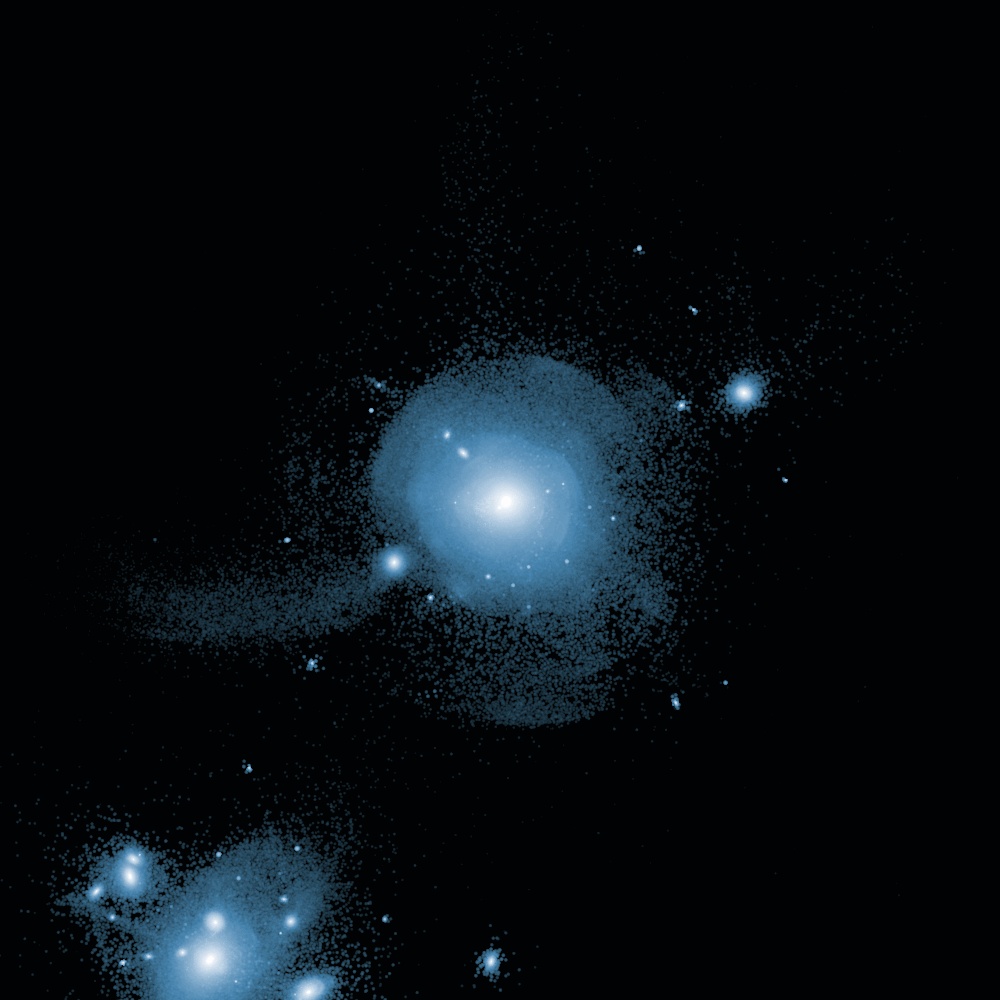}\llap{\makebox[\wd1][l]{\raisebox{0.8\wd1}{\includegraphics[width=0.28\textwidth]{LA9_Label-eps-converted-to.pdf}}}}
      \includegraphics[scale=0.23]{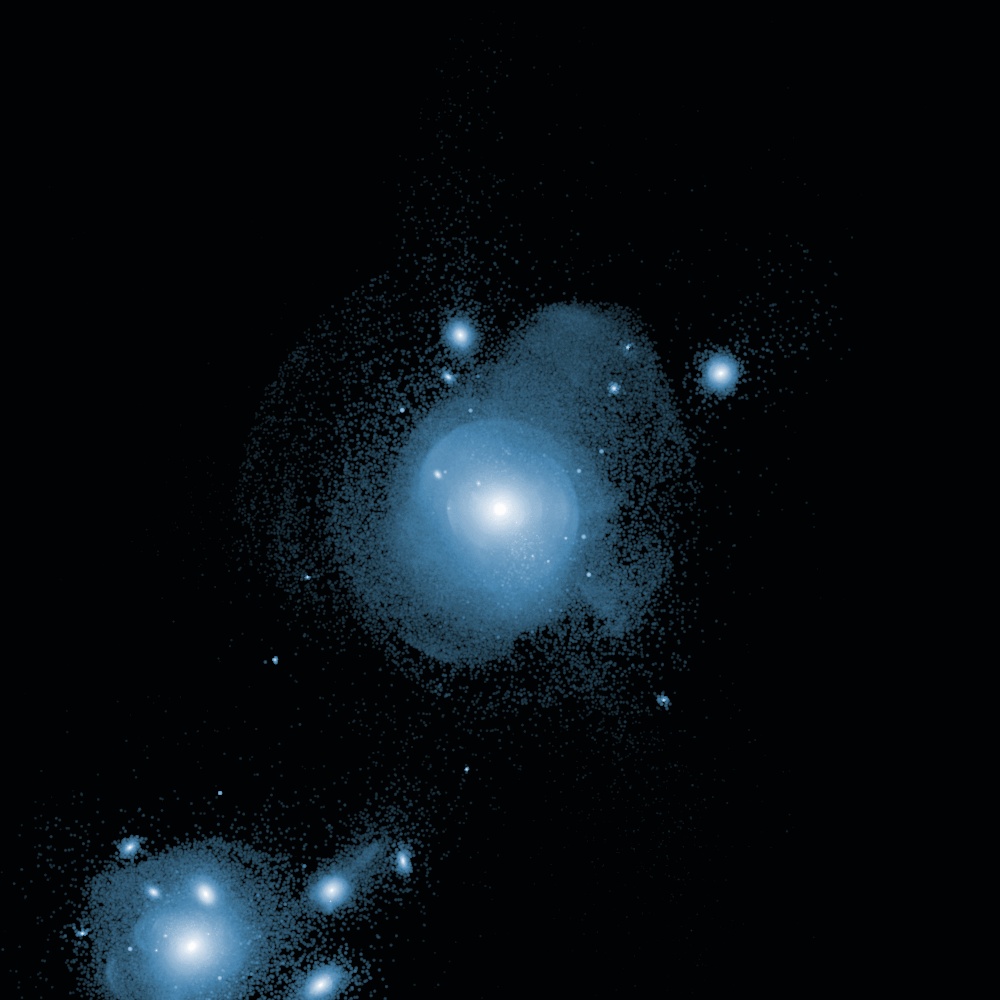}\llap{\makebox[\wd1][l]{\raisebox{0.8\wd1}{\includegraphics[width=0.28\textwidth]{LA11_Label-eps-converted-to.pdf}}}} \\
    
    \caption{The projected $z=0$ stellar mass density in the four high-res simulations. As in Fig.~\ref{fig:DM_Images}, the four panels are AP-HR-CDM (top left), AP-HR-SIDM10-Rec9 (top right), AP-HR-LA9 (bottom left) and AP-HR-LA11 (bottom right). The image intensity indicates the projected density; unlike Fig.~\ref{fig:DM_Images} no velocity dispersion information is encoded. Each image is 1.2~Mpc on a side, and the density is projected over a distance of 0.8~Mpc both in front of and behind the image centre.}
    \label{fig:Star_Images}
\end{figure*}

We preview the results of these simulations in Figs.~\ref{fig:DM_Images} and~\ref{fig:Star_Images}, in which we present a rendering of the dark matter and stars respectively in the four high resolution simulations that were run to completion at $z=0$. We focus on the more massive of the two galaxies; that is, the M31 analogue. Each rendering is 1.2~Mpc in width and 0.8~Mpc deep either side of the M31 analogue centre-of-potential. In the dark matter panels we present the projected density using the image intensity and the density-weighted velocity dispersion using the hue, where purple indicates the lowest velocity dispersions and yellow the highest.

 The large-scale structure of the dark matter surrounding the CDM version of the halo is reproduced almost exactly in the other three versions.  This is reassuring for two reasons. First, one requirement of any alternative dark matter model is to preserve the successes of CDM at large scales. Second, it is much easier to compare subhalos on a halo-by-halo basis when the large-scale structure in the given simulation is invariant with changes to the dark matter model. Previous work has shown that WDM models much warmer than those used in this study can change how the M31 and MW analogue haloes evolve significantly relative to the original CDM halo \citep{Libeskind13}. The existence and location of the largest dwarf haloes is largely preserved, although changes in both can often be attributed to stochastic effects and are therefore not necessarily an expression of the difference in the underlying physics. We recover the familiar result that the abundance of small structures in  the two WDM models is suppressed relative to CDM, and more so in LA11 than in LA9. 

The differences between CDM and the SIDM10 model are more subtle. There is a hint of the suppression of subhaloes in the halo centre due to interactions between the subhalo and the host as found by e.g.~\citet[][]{Vogelsberger12}. There is also some visual evidence that the centre of the M31 analogue halo is rounder in SIDM10 than it is in CDM, which is a known output of isotropic scattering from dark matter self-interactions  \citep{Colin02,Peter13,Tulin17} and will be followed up in future work.

 The differences in the dark matter distribution are also expressed in the stellar density fields (Fig.~\ref{fig:Star_Images}). The WDM models have lower numbers of low-mass galaxies relative to CDM, reflecting the suppression in the number of dwarf haloes in these models. The abundance of dwarf galaxies in the SIDM10 run is largely similar to CDM except for the absence of haloes in the M31 analogue centre and the apparently enhanced destruction rate of some larger haloes. These results are consistent with the evaporation of SIDM10 haloes due to interactions between satellite and host halo particles, although for the disruption of larger haloes the cored density profiles, different galaxy formation model, and stochastic effects could well be more significant factors \citep{Dooley16}.

The spatial extent of the stellar haloes is apparently larger in the sterile neutrino runs, and further investigation is required to show whether this is a product of the model or instead a stochastic or numerical feature. It is possible that the caustics in the WDM stellar haloes are more pronounced than in their CDM counterparts, and it will be interesting to determine first whether or not such an effect exists, and if so, show whether the changes to the abundance and structure of WDM haloes are relevant for this observable; we will address this question in future work. Finally, we note that comparing each stellar density panel with its dark matter counterpart reveals which dark matter haloes host galaxies. Remarkably, it is very difficult to predict from the size of each dark matter halo alone whether it will host any stars, thus emphasizing the importance of using a high quality galaxy formation model to populate haloes with galaxies.

 We end this section with a discussion of the observations that we compare to in this study. We adopt the compilation of observed stellar masses and age distributions of \citet{Digby19} and extract the positions of those galaxies from \citep{McConnachie12}. We restrict our comparison to those observed galaxies in which the main sequence turnoff (oMSTO) has been detected, because without such a detection the systematic uncertainties in stellar population age become exceedingly large, especially when we are primarily interested in the oldest stellar populations. 

\begin{table*}
    \centering
      \caption{Properties of the simulations used in this paper: simulation name, dark matter particle mass ($m_\rmn{DM}$), gravitational softening length ($\epsilon$), dark matter model, galaxy formation model, redshift of reionization ($z_\rmn{re}$), redshift at which the simulation ends ($z_\rmn{end}$), and the original reference.}
      \centering
    \begin{tabular}{l|c|c|c|c|c|c|l|}
    \hline
        Run & $m_\rmn{DM}$~$(\msun)$& $\epsilon$~(kpc) & DM model & GF model & $z_\rmn{re}$ &$z_\rmn{end}$ & Ref. \\
\hline
    AP-HR-CDM(-Ref) &  $5\times10^{4}$ & 0.13  & CDM & Ref & 11.5 & 0 & \citet{Sawala16a}\\
    AP-HR-CDM-Rec9 & $5\times10^{4}$ & 0.13  & CDM & Rec & 9 & 5.2 & This work \\
    AP-MR-CDM-Ref7 & $6\times10^{5}$ & 0.35  & CDM & Ref & 7 & 0 & Fattahi et al. (in prep.) \\
    AP-MR-CDM(-Ref, 1-6) & $6\times10^{5}$ & 0.35  & CDM & Ref & 0 & 0 & \citet{Sawala16a}\\    
    &&&&&&&\\
    AP-HR-LA9(-Ref) &  $5\times10^{4}$ & 0.13  & $m=7$~keV, $L_{6}=9$ & Ref & 11.5 & 0 & This work\\
    AP-HR-LA11(-Ref) &  $5\times10^{4}$ & 0.13  & $m=7$~keV, $L_{6}=11.2$ & Ref & 11.5 & 0 & \citet{Lovell19a}\\
    AP-MR-LA11(-Ref) & $  6\times10^{5}$ & 0.35  & $m=7$~keV, $L_{6}=11.2$  & Ref & 11.5 & 0 & This work\\
     AP-MR-L10(-Ref, 1-6) & $  6\times10^{5}$ & 0.35  & $m=7$~keV, $L_{6}=10$ & Ref & 11.5 & 0 & \cite{Lovell17b}\\
     &&&&&&&\\
    AP-HR-SIDM10-Rec9 & $5\times10^{4}$ & 0.13  & SIDM(10~$\rmn{cm}^2\rmn{g}^{-1}$) & Rec & 9 & 0 & This work\\
    AP-MR-SIDM10-Rec9 & $6\times10^{5}$ & 0.35  & SIDM(10~$\rmn{cm}^2\rmn{g}^{-1}$) & Rec & 9 & 0 & This work \\
 \hline
    \end{tabular}
  
    \label{tab:sims}
\end{table*}

 \section{Results}
 \label{res}
 
 We begin the presentation of our results with a resolution study of the stellar mass at $z=0$, the total halo mass including all mass species -- gas, dark matter, black holes, and stars combined --  at $z=0$, and stellar mass formed at $z>6$, using the CDM, LA11 and SIDM10 versions of volume AP-1. We then proceed to a compare the models to one another, first for the halo and stellar masses, then the star formation histories and the stellar population formed within the first gigayear, and ending with a discussion of the oldest stellar populations extant at $z=0$. Our study of the first gigayear stellar populations also features a comparison to the \citet{Digby19} collection of observations.
 
 \subsection{Resolution}
 
\begin{figure*}
     \centering
     \includegraphics[scale=0.34]{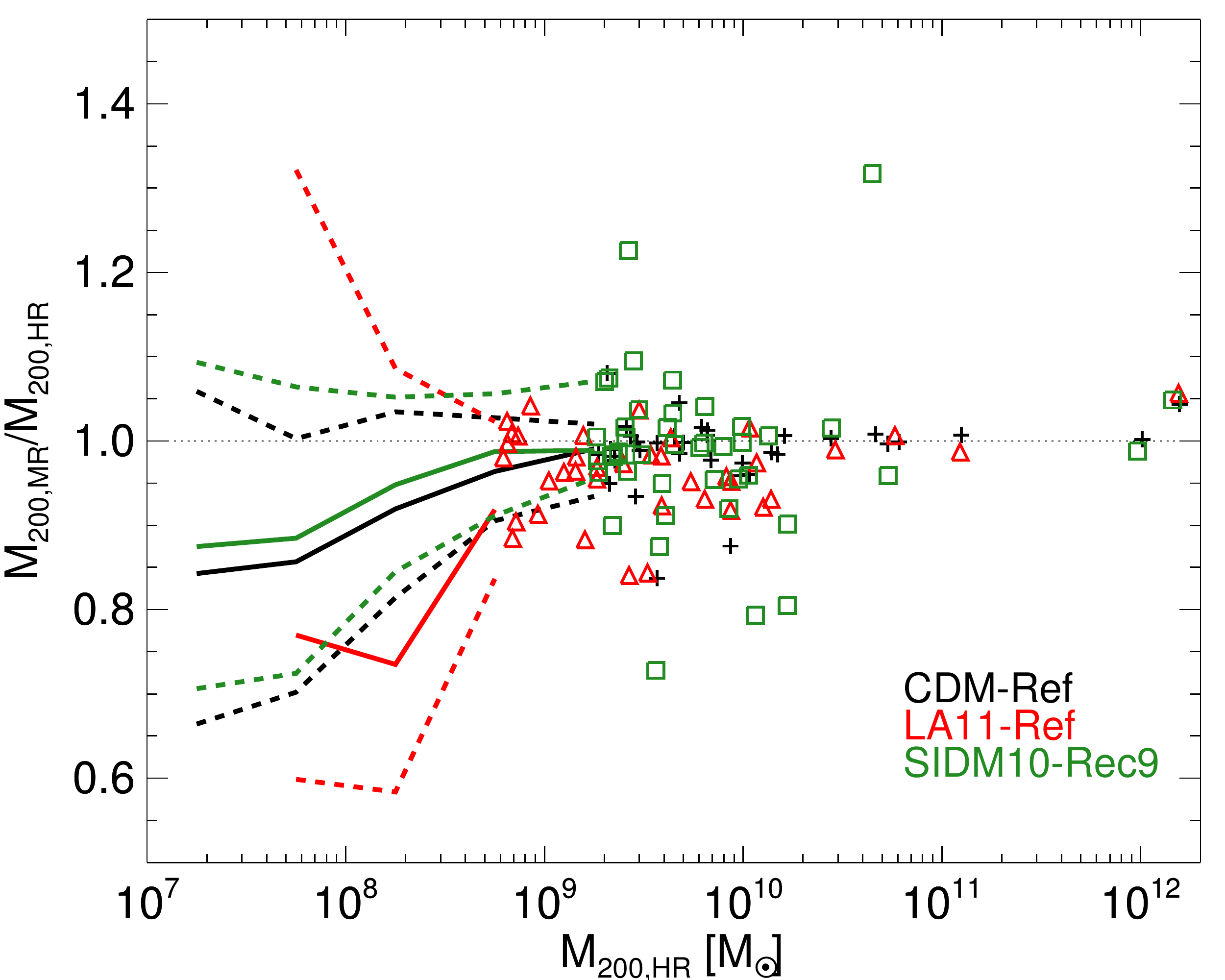}
      \includegraphics[scale=0.34]{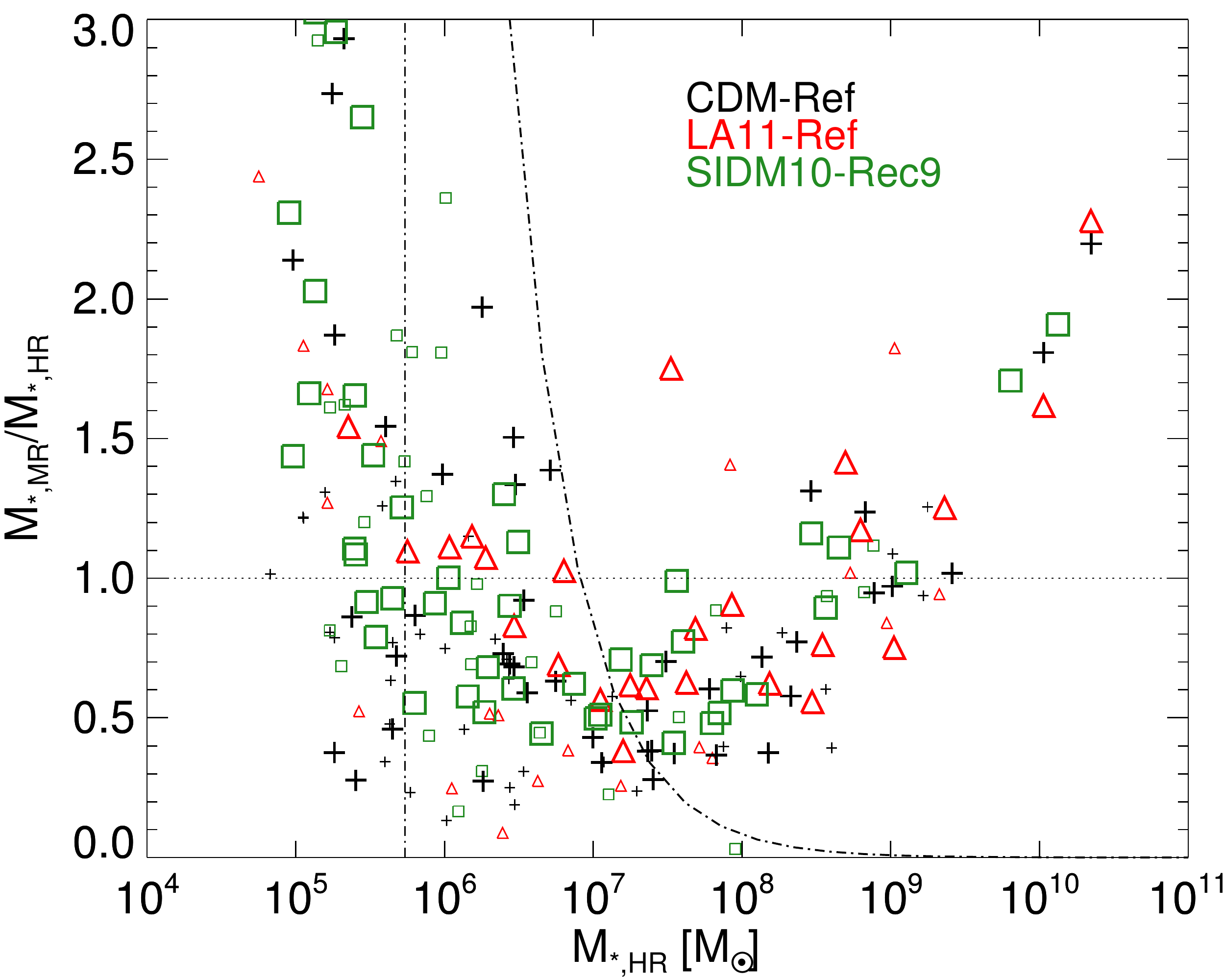}
     \caption{Resolution dependence of halo mass (left-hand panel) for isolated haloes and stellar mass (right-hand panel) for both isolated haloes and subhaloes. MR galaxy halo and stellar masses are shown as as a function of, and normalised by, the same quantity for the HR-matched galaxy. CDM galaxies are shown as black plus signs, LA11 galaxies as red triangles and SIDM10 galaxies as green squares. The two dot-dashed lines in the right-hand panel indicate the stellar masses at which the galaxies have 100 star particles, which is the criterion for resolving stellar mass adopted by \citet{Schaye15}: the vertical line corresponds to the HR limit and the curved line to the MR limit. In the right hand panel we show the results for isolated haloes as large symbols and for subhaloes/satellites as small symbols. In the left-hand panel, the data are shown below $3\times10^{9}\msun$ ($1\times10^{9}\msun$) for CDM and SIDM10 (LA11) as a median relation (solid line) with the 68~per~cent scatter regions (dashed lines).}
     \label{fig:MsRes}
 \end{figure*}

 The halo mass and stellar mass of a galaxy are two of its most basic properties, and therefore influence how we interpret the results of its star formation history. The former has been shown to converge well with improved mass and spatial resolution in $N$-body simulations \citep{Springel08b}; performing the same exercise with stellar mass has been complicated by the change in particle number, and thus our halo matching method comes into its own. We plot the relationship between the MR and HR versions of the halo mass and stellar mass for CDM, LA11 and SIDM10 haloes in Fig.~\ref{fig:MsRes}. Halo mass is defined as the $z=0$ mass enclosed within a radius of mean enclosed density 200 times the critical density for collapse, $M_{200}$, which we refer to as the virial mass. Our choice of stellar mass, $M_{*}$, is the total mass of stars within a spherical 30~kpc aperture of the galaxy/halo centre-of-potential. In the halo mass plot (left-hand panel) we include all isolated haloes independently of whether or not they are luminous: fewer than 0.1~per~cent of SIDM10 haloes in the mass range $[10^{7},10^{8}]\msun$ are luminous, compared to 0.2~per~cent of CDM haloes and 0.8~per~cent of LA11 haloes. For the $M_{*}$ plot (right-hand panel) we only include haloes that are luminous in both the MR an HR copies of each halo, and include subhaloes as well as isolated haloes. \footnote{ We note that the mass resolution is sufficient in all of our simulations to resolve the peak mass of star-forming haloes, $>10^8\msun$, with an accuracy better than 10~per~cent \citep[c.f.][]{Springel08b,Ludlow19a} and therefore we resolve all haloes that could plausibly form a galaxy; the absence of galaxies in subhaloes is therefore due to the interaction between mass / spatial resolution and the astrophysics model.} 
 
 We recover halo mass and stellar mass relationships between the MR and HR simulations that are remarkably similar for all three models, with the possible exception of the lower masses of MR LA11 galaxies with $M_\rmn{200,HR}<10^{9}\msun$. The masses of HR CDM $M_\rmn{200,HR}>10^{9}\msun$ haloes vary on average by 8~per~cent with respect to their MR counterparts. However, the median ratio between the matched pair halo masses is 0.99: there is no systematic change in halo mass between resolutions, and we expect that much of the variation is driven by stochastic factors. The behaviour of stellar mass is more complicated: in all three cases MR systematically forms more stars at $M_\rmn{*,HR}>10^{9}\msun$, whereas the opposite is the case at $M_\rmn{*,HR}=[10^{7}-10^{8}]\msun$ \citep{Ludlow19b}. Note that below $10^{7}\msun$ the MR galaxies have fewer than 100 star particles; \citet{Schaye15} showed that resolution-related sampling effects caused EAGLE galaxies with fewer than 100 baryonic particles to gain stellar mass spuriously. Finally, we note that there is some evidence that subhaloes have lower MR stellar masses at fixed HR stellar mass $\sim10^{6.3}\msun$ than do isolated haloes, possibly indicating that low-resolution haloes are more easily stripped of material by tidal forces than their high-resolution counterparts.
 
\begin{figure}
     \centering
     \includegraphics[scale=0.34]{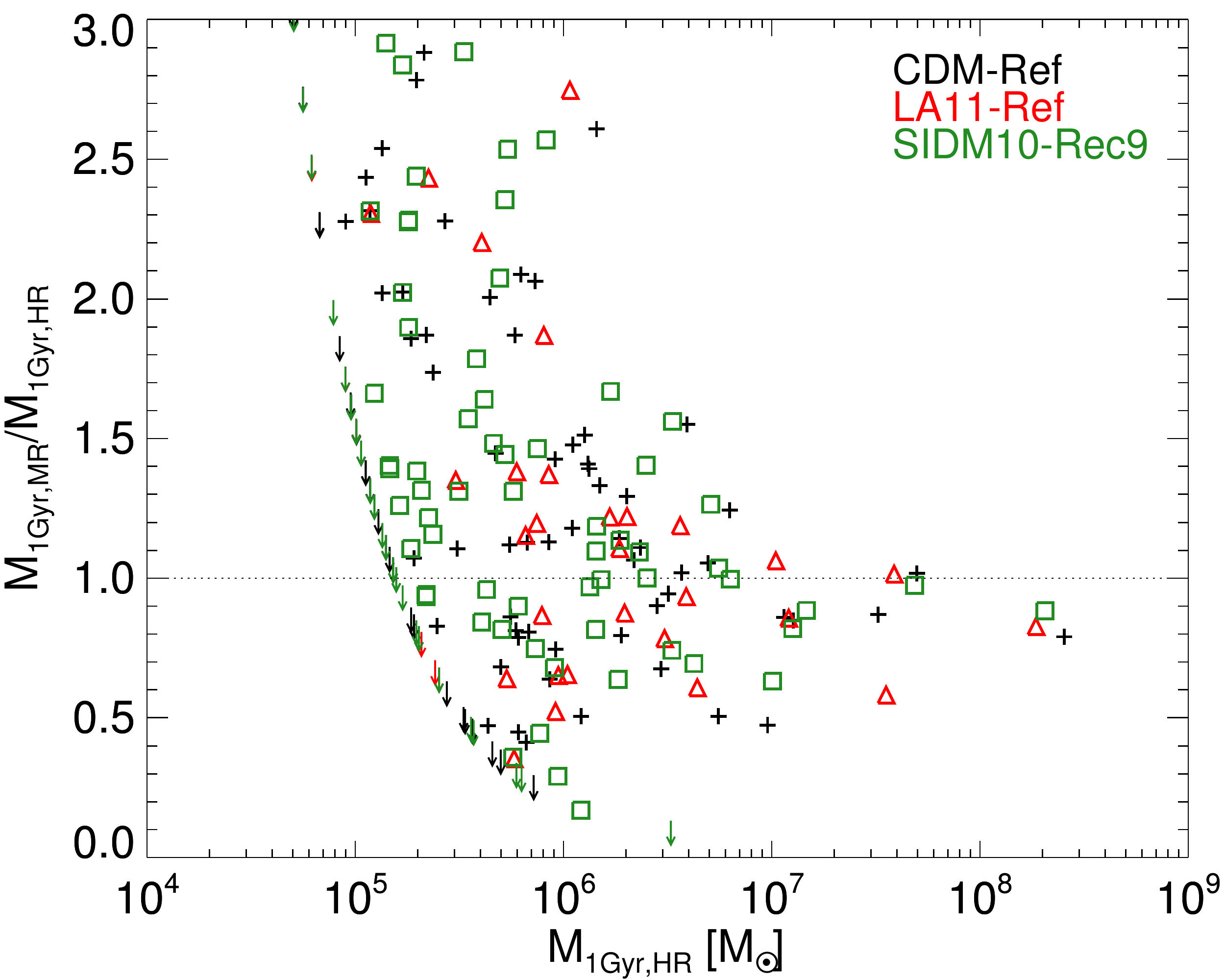}
     \caption{The $z=0$ stellar mass formed in MR galaxies in the first gigayear after the Big Bang, normalised by the same quantity computed for the counterpart HR galaxy. CDM galaxies are shown as black plus signs, SIDM10 galaxies as green squares and LA11 galaxies as red triangles. MR galaxies that contain no stars formed in the first gigayear are shown as arrows, and follow the curve that corresponds to the stellar particle mass resolution for the MR simulation ($1.5\times10^5\msun$). We include both haloes and subhaloes.}
     \label{fig:G1MRes}
 \end{figure}

The stellar mass formed within the first gigayear of the Universe is accessible to observations, albeit with large systematic uncertainties, and likely to be affected by the nature of dark matter. For each MR-HR galaxy pair we compute the total stellar mass, again within the 30~kpc spherical aperture, at $z=0$ that was formed less than 1~Gyr after the Big Bang, $M_\rmn{1Gyr}$, and plot the MR $M_\rmn{1Gyr}$ normalised by HR $M_\rmn{1Gyr}$ in Fig.~\ref{fig:G1MRes}.
 
The scatter in this plot for CDM and LA11 is slightly larger than for the total stellar mass shown in Fig.~\ref{fig:MsRes}. For HR $M_\rmn{1Gyr}>10^{6}\msun$ the average difference between HR and MR is 60~per~cent, and the median ratio between the counterparts is 1.09, so MR is biased to generating more stellar mass in the first gigayear relative to HR. For HR $M_\rmn{1Gyr}<10^{6}\msun$ many of the MR counterparts lack star particles due to shot noise. We conclude that our HR results for the different models are therefore subject to uncertainty from resolution effects, although it is reassuring that the systematic offset over this factor of 8 in mass resolution is only 9~per~cent.
 
To summarize, we have shown that the bound halo mass varies by about 10~per~cent between HR and MR realisations, although the systematic bias is of the order of $<5$~per~cent. Therefore the average halo mass is converged even though the variations for individual haloes are often of the order of tens of per~cent. By contrast, the total stellar mass changes significantly with resolution: massive galaxies form over twice as many stars in MR as HR, while the opposite is true for dwarf galaxies. The reason for this lack of convergence follows from the fact that the subgrid models of EAGLE were calibrated at EAGLE-Ref resolution, or, in the case of AP-HR-SIDM10-Rec9, at EAGLE-Rec resolution. It is therefore not surprising that the use of this calibration in simulations with 10 or 100 times better resolution result in a lack of convergence in the stellar mass. 
The stellar mass formed within the first gigayear does not shown any systematic trend above the statistical uncertainties; we also showed that the resolution dependence is the same for CDM, LA11, and SIDM10, even though SIDM10 was run with the Rec9 model rather than Ref.
 
 \subsection{Dark matter model}

 \subsubsection{Halo and stellar masses}

\begin{figure*}
     \centering
      \includegraphics[scale=0.34]{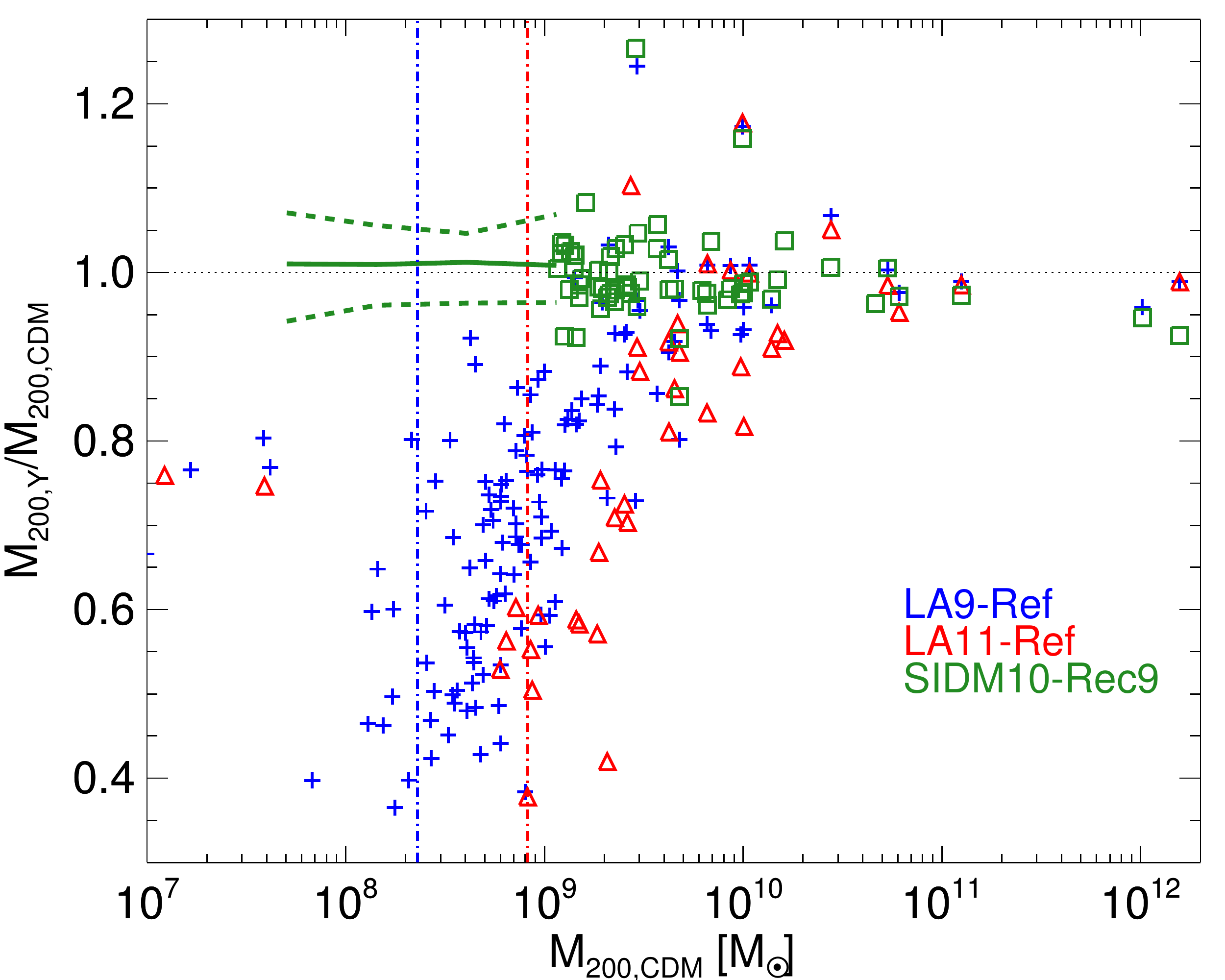}
     \includegraphics[scale=0.34]{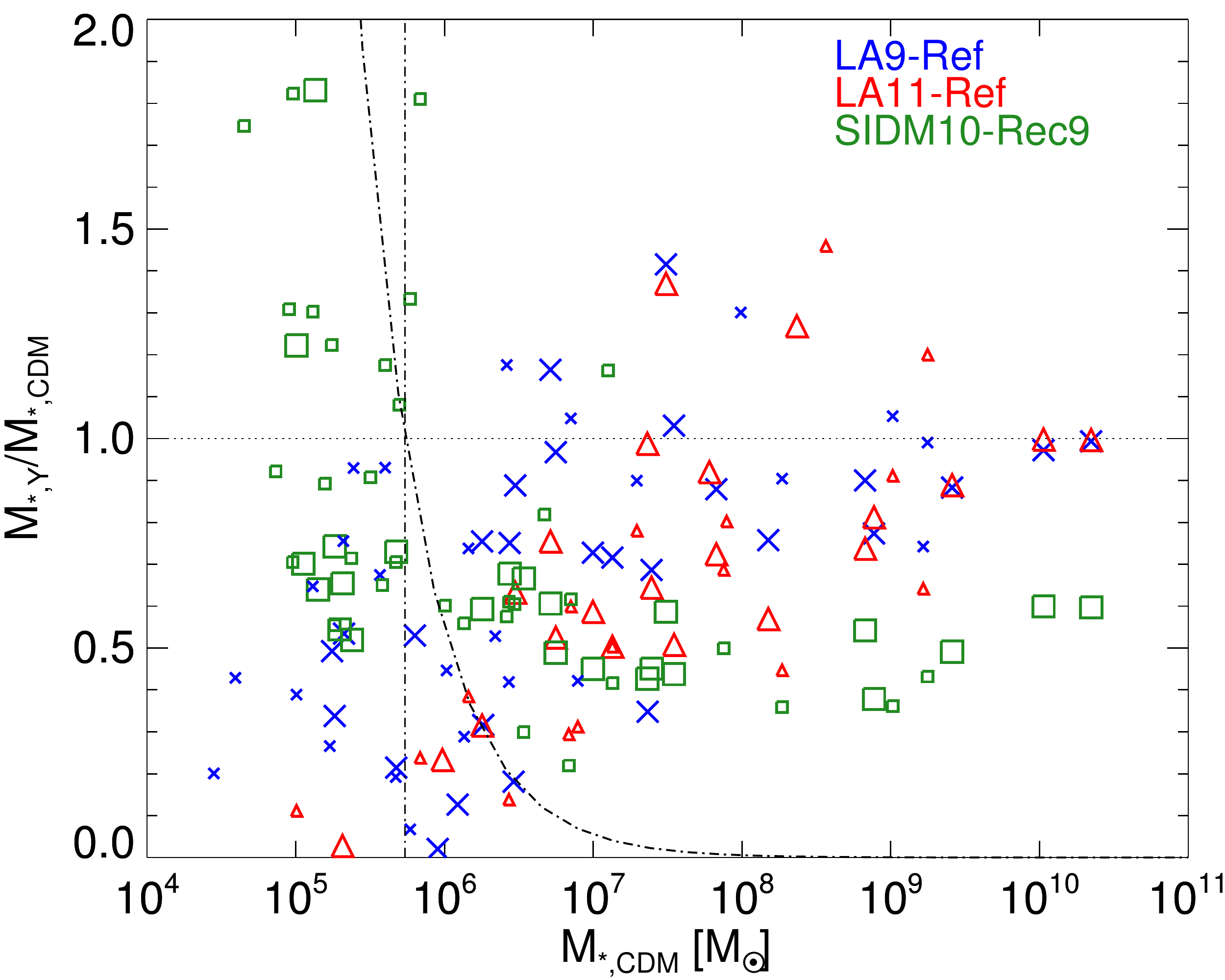}
     \caption{Halo mass (isolated haloes, left-hand panel) and stellar mass (isolated haloes and subhaloes, right-hand panel) of WDM and SIDM galaxies normalised by the same quantity for the CDM-matched galaxy. Matches to LA11 are shown as red triangles, LA9 galaxies as blue crosses and SIDM10 galaxies as green squares. In all three cases the models are matched to the CDM-Ref simulation. As in Fig.~\ref{fig:MsRes}, the SIDM10 data are shown as a median line below $10^{9}\msun$ (solid line) with the 68~per~cent scatter regions (dashed lines). We add the half-mode mass for the two WDM models as dot-dashed lines. In the right-hand panel the 100 star particle resolution limits are shown with dot-dashed lines, and isolated haloes (subhaloes) are shown as large (small) symbols.}
     \label{fig:MsMod}
 \end{figure*}

Having shown the degree to which the halo mass, the total stellar mass, and the stellar mass formed in the first gigayear converge with resolution, we consider how these quantities change in the case of the WDM and SIDM models relative to CDM. We start with the halo and stellar masses of LA9/LA11/SIDM10 galaxies, as matched to CDM counterparts, see Fig.~\ref{fig:MsMod}. All of the simulations used in this section are HR.
 
  SIDM10 agrees with CDM over the whole mass range shown, with 73~per~cent of haloes agreeing to better than 10~per~cent. This result is expected because the SIDM model preserves halo masses as long as evaporation is insignificant\footnote{ We conjectured in Figs.~\ref{fig:DM_Images} and \ref{fig:Star_Images} that the satellite abundance is suppressed by evaporation. We therefore expect that for SIDM10 evaporation is a rapid process in which a halo that initially has a CDM-like mass is quickly destroyed, thus keeping the halo mass the same for largely unstripped haloes and suppressing subhalo abundances simultaneously. We leave investigation of this effect to future work.}. The sterile neutrino models instead exhibit suppressed halo masses relative to CDM below $\lsim3\times10^{9}\msun$ ($\lsim10^{10}\msun$) for LA9 (LA11), as anticipated by \citet{Bozek19} and \citet{Lovell19b}. The mass at which the suppression is a factor of two occurs up to 50~per~cent higher than the half-mode mass, defined as the mass that corresponds to the wavenumber where the square root of the WDM-CDM power spectrum ratio is 0.5, and is thus slightly more pronounced than found for the ETHOS model at high redshift in \citet{Lovell19a}. This suppression in mass is mirrored in the stellar masses, for which CDM and the WDM models are roughly similar -- albeit it with $\sim20$~per~cent scatter -- above $\sim10^{8}~\msun$. Towards lower masses the WDM galaxies become progressively less massive than their CDM counterparts. The LA11 stellar masses start to be suppressed at the CDM stellar mass of $10^8\msun$, and LA9 similarly diverges at $3\times10^{7}\msun$. By contrast, the SIDM10 simulation shows a marked suppression in stellar mass at CDM galaxies of $M_{*}>10^{7}\msun$, typically by a factor of 2. Given that the halo mass function is the same in SIDM10 as in CDM, we conjecture that this suppression is due to a combination of the lower densities in SIDM10 haloes and also the stronger feedback given the use of Rec rather than Ref parameters, which together lead to lower star formation rates at late times. We find that the median stellar mass fraction of SIDM10-Rec9 haloes at $z=0$ is 40~per~cent that of their CDM counterparts. Finally, we do not find any dramatic evidence of a difference in the stellar mass changes between satellites and isolated haloes. 
  
  \subsubsection{Star formation histories}
 
 The sensitivity of the star formation histories of galaxies to the nature of dark matter models has been described at length in the literature, especially for WDM models. Delays to the onset of star formation -- and correspondingly lower average stellar ages -- have been demonstrated for isolated galaxies \citep{Bozek19,Maccio19}, and for the ensemble of LG dwarfs at APOSTLE MR resolution \citep{Lovell17b}. A similar delay has been shown in high redshift simulations of WDM \citep{Liu19} and for the ETHOS model, which, similar to WDM, features a cut-off in the matter power spectrum \citep{Lovell19b}\footnote{ETHOS also features dark matter self-interactions. We expect that the delay in formation times in that specific model is due to the cut-off rather than those self-interactions, and we show this implicitly in this study when comparing the SIDM10 and WDM models.}. Here we take advantage of the combination of high resolution, plus the environment of the LG-analogue, to ascertain the behaviour of these models as they would apply to the LG. We will compare and contrast the star formation histories of a population of galaxies that are present in all four of our HR simulations, then infer when the oldest extant LG dwarf galaxy stars formed, and provide a brief comparison to observations. 
 
\begin{figure*}
     \centering
     \includegraphics[scale=0.69]{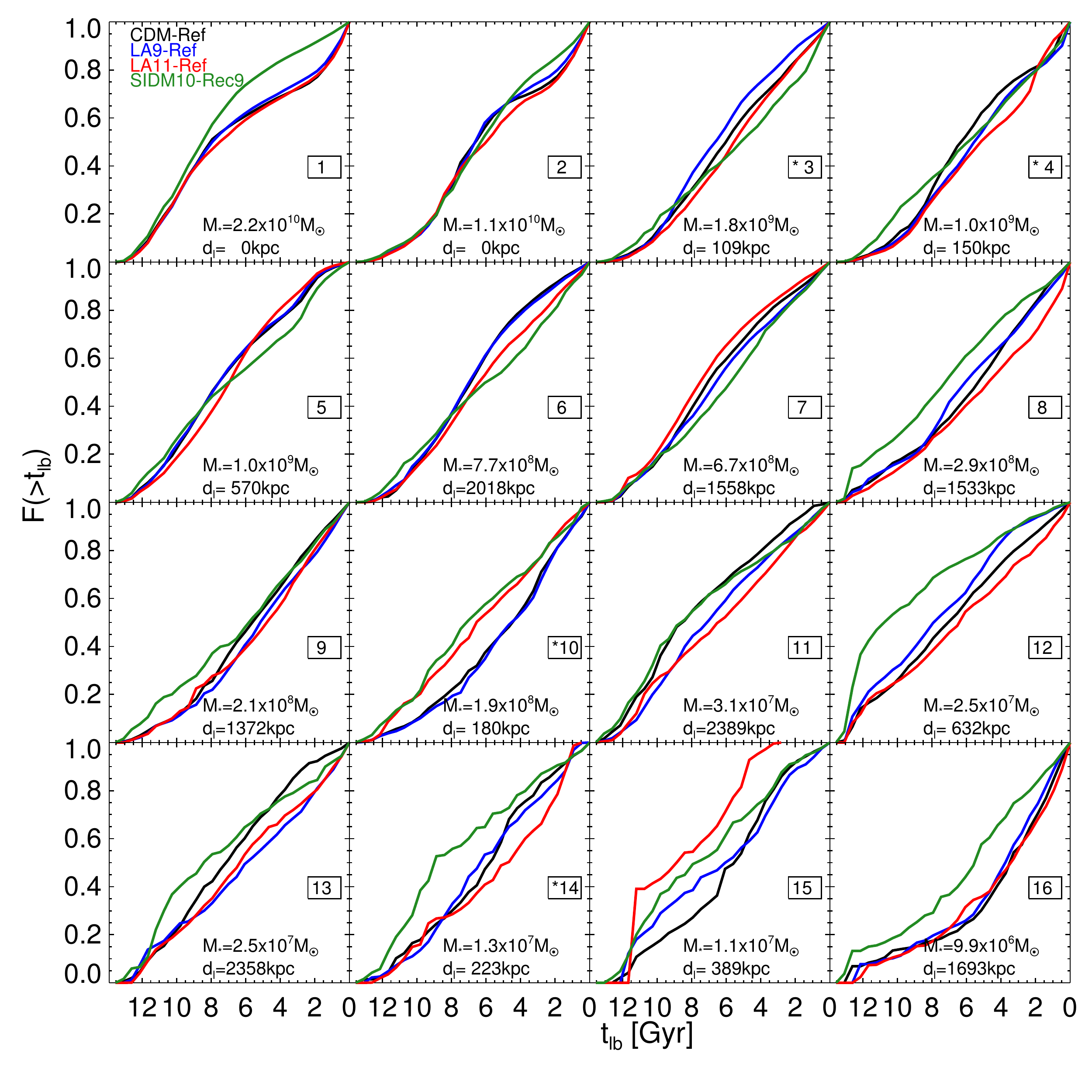}
     \caption{Cumulative star formation histories of 16 galaxies matched between all four high resolution simulations, normalised to the $z=0$ stellar mass of each galaxy. CDM is shown in black, LA9 in blue, LA11 in red, and SIDM10 in green. The galaxies are ordered by total CDM $z=0$ stellar mass from left to right and top to bottom; the first two galaxies are the M31 and MW analogues. The quoted distance for satellites is the distance to the host galaxy, and for field galaxies is the distance to the M31-MW system barycentre; satellites are marked with an asterisk. Both the stellar mass and distance quoted are for the CDM version of each galaxy.}
     \label{fig:SFH_AM}
 \end{figure*}

 We select 16 CDM galaxies that are located within 2.5~Mpc of the MW-M31-analogue barycentre and have high fidelity ($R\ge0.9$) matches in all three of the alternative dark matter simulations. We plot the cumulative star formation histories for these four models in Fig.~\ref{fig:SFH_AM}. Here we normalise each star formation history by each galaxy's $z=0$ mass; we will consider an alternative normalisation -- to the counterpart CDM galaxy stellar mass -- in a subsequent plot. Of these 16, two are the M31 and MW-analogues, 4 are satellites, and the remainder are field galaxies. All of the CDM versions have $M_{*}>10^{7}\msun$, which is of the order of the stellar mass of the Leo~I dSph \citep{McConnachie12}.     
 
 There is evidence of a systematic difference between the CDM and WDM haloes in these star formation histories. It is common for LA9 to track CDM more closely than LA11 (e.g. panels 5, 6, 8, and 10); the divergence of LA11, typically at later times, is consistent with the power spectrum cut-off delaying structure formation, but it occasionally happens earlier (panels 7, 10, 15) at least in terms of its own formation time. The SIDM10 galaxies tend to form most of their stellar mass earlier than their CDM counterparts. At present, it is likely that this delay is due to the different galaxy formation model rather than the role of the self-interactions, a question that we leave to future work.
 
\begin{figure*}
     \centering
     \includegraphics[scale=0.68]{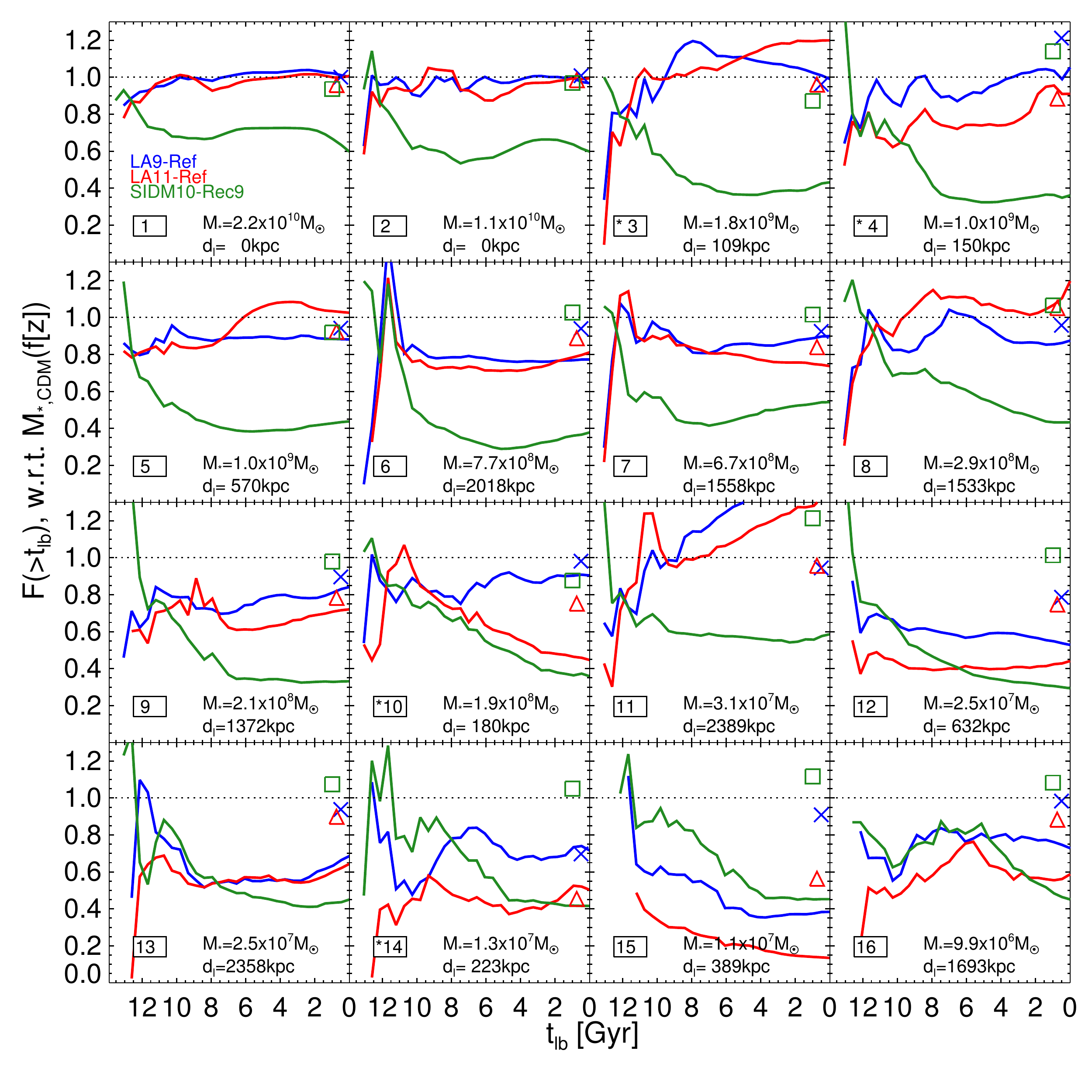}
     \caption{Cumulative star formation histories of the galaxies from Fig.~\ref{fig:SFH_AM} divided by the star formation history of the counterpart CDM-simulation galaxy. The coloured crosses indicate the approximate ratio of the halo masses of the three runs to their CDM counterpart at $z=0$, as given by the square of the ratio of their $V_\rmn{max}$.}
     \label{fig:SFH_AMdC}
 \end{figure*}

The star formation histories of the WDM models show a large degree of variation relative to CDM, but we have also shown that both the $z=0$ halo mass and, particularly, stellar mass are suppressed in the presence of a matter power spectrum cut-off (Fig.~\ref{fig:MsMod}). We correct for the effect of this halo mass suppression by re-normalising our alternative dark matter star formation histories by their CDM counterparts, so that we divide each $M_{*}(z)$ function by the corresponding CDM $M_{*}(z)$. We plot the results for our subsample of 16 galaxies in Fig.~\ref{fig:SFH_AMdC}.  
 
 There is a clear suppression of the SIDM10 halo star formation histories relative to CDM that is common to all 16 galaxies at $t_\rmn{lb}<12$~Gyr. For the WDM models the suppression becomes stronger towards lower stellar masses, and is stronger for LA11 than LA9, although not without exception (panels 3, 5, and 8). Panel 11 bucks the trend with higher stellar masses than the CDM counterpart: in this case the CDM galaxy slows down its star formation rate after $t_\rmn{lb}\sim8$~Gyr but the WDM galaxies continue to grow at pace. We noted above that both the $z=0$ stellar and total halo masses are suppressed in the WDM models relative to CDM, and therefore it is reasonable to hypothesise that the two scale together. We therefore plot the ratio of the final halo masses on each panel as crosses. In some cases this scaling is approximately accurate for LA11 and LA9 (panels 4, 8, 9,14), but it is common for the stellar mass to be suppressed by up to 50~per~cent for LA11 as compared to mostly less than 20~per~cent for the halo mass. The median change in baryon fraction from CDM to LA9 is 18~per~cent, and from CDM to LA11 is 28~per~cent. It is therefore likely that further effects are at play: possibilities include the evaporation of gas by reionization before the halo collapses, and the weaker potential of the less massive WDM haloes making the removal of gas by feedback and/or interactions with gas in the cosmic web more efficient \citep{BenitezLlambay13}.   
 
\begin{figure}
     \centering
     \includegraphics[scale=0.34]{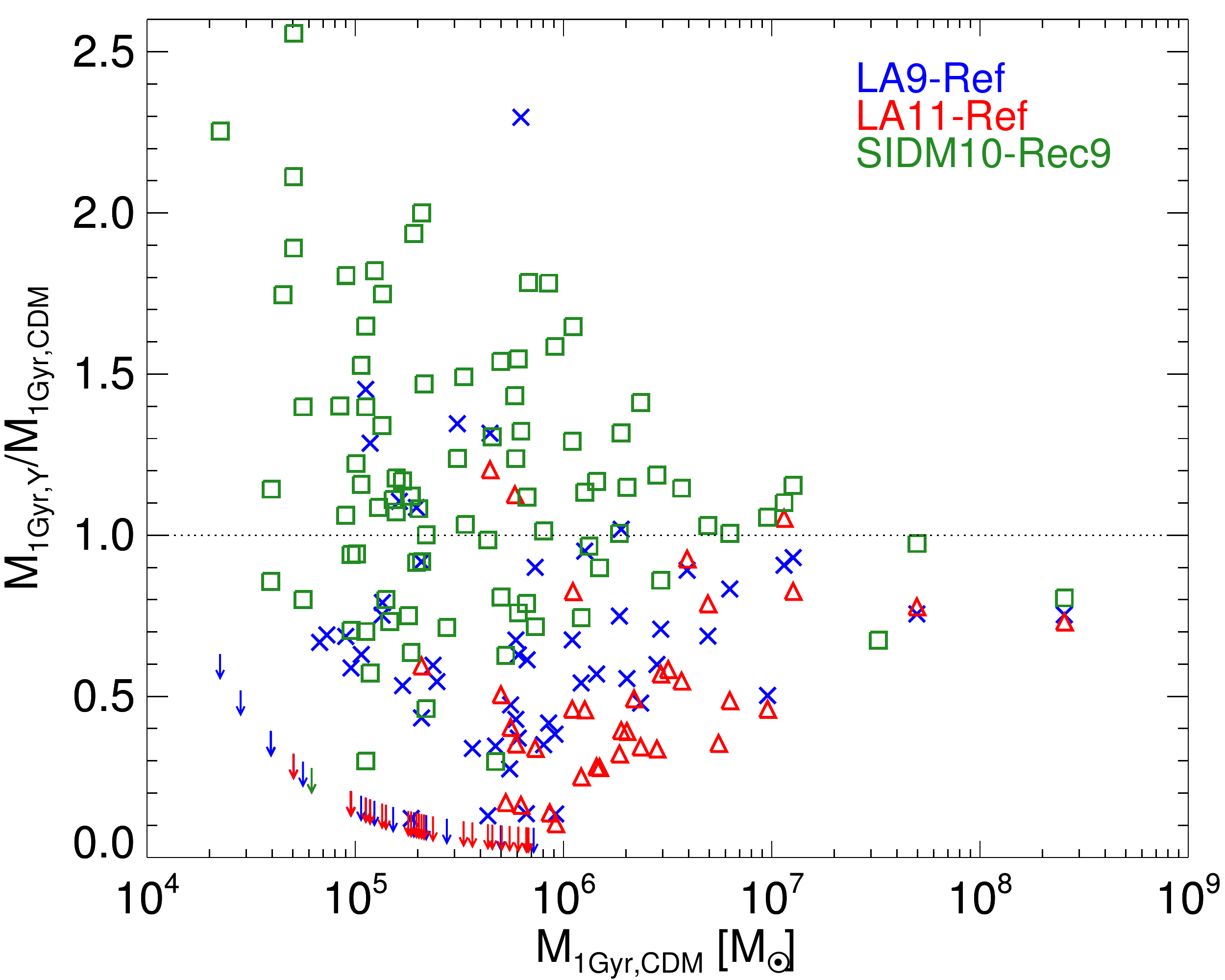}
     \caption{The stellar mass formed in WDM, SIDM10 and CDM galaxies in the first gigayear after the Big Bang, as normalised to the CDM counterpart. LA9 galaxies are shown as blue crosses, LA11 galaxies as red triangles and SIDM10 galaxies as green squares. In all three cases the galaxies are matched to the CDM-Ref simulations. Galaxies that contain no stars formed in the first gigayear in the WDM and SIDM10 runs are shown as arrows, which track the one-star-particle curve, i.e. $M_\rmn{1Gyr,Y}=1.25\times10^{4}\msun$. Conversely, there is a small number of haloes that form stars in WDM and/or SIDM10 but not in CDM and these are not shown in this plot: they include two LA9 haloes, one LA11 halo, and fourteen SIDM10 haloes.}
     \label{fig:G1MMod}
 \end{figure}

 One of the features common to many of the LA9 and LA11 galaxies presented here, and to a lesser degree the SIDM10 counterparts, is a sharp increase in stellar mass at very early times, $t_\rmn{lb}>11$~Gyr for galaxies 3, 6, 7, 8, 10, 13. This time corresponds to the starburst associated with the monolithic collapse of the first star-forming structures due to the power spectrum cut-off \citep{Bose16c,Lovell19b}. However, although the starburst produces a larger quantity of stars in a short period of time, it is not sustained long enough to form more stars overall than happens in CDM, and so the total stellar mass formed within the first gigayear is suppressed relative to CDM. The cumulative stellar mass formed by this time is accessible to observations \citep{Digby19}, and we therefore consider the stellar mass formed within the first gigayear after the Big Bang, plotting the WDM/SIDM10-CDM ratios of $M_\rmn{1Gyr}$ as a function of CDM $M_\rmn{1Gyr}$ in Fig.~\ref{fig:G1MMod}. 
 
 The stellar mass formed within the first gigayear is significantly suppressed in WDM, and more so for LA11 than LA9. This is the case even for galaxies in which the total (i.e. $z=0$) stellar mass is the same in the two models, around $10^{9}\msun$. There are large populations of galaxies in which the WDM counterpart does not form any stars in the first gigayear whereas the CDM case does; the converse is not true, in that there are  only two LA9 galaxies and one LA11 galaxy (not visible in the plot), that start forming stars this early in WDM but not in CDM. The SIDM10 model instead is similar to CDM, even exceeding its $M_\rmn{1Gyr}$ below $10^{7}\msun$. This is most likely due to the later redshift of reionization assumed in this run enabling star formation to take place earlier, as we will demonstrate in Section~\ref{sec:oosf}.
 
\begin{figure*}
     \centering
     \includegraphics[scale=0.68]{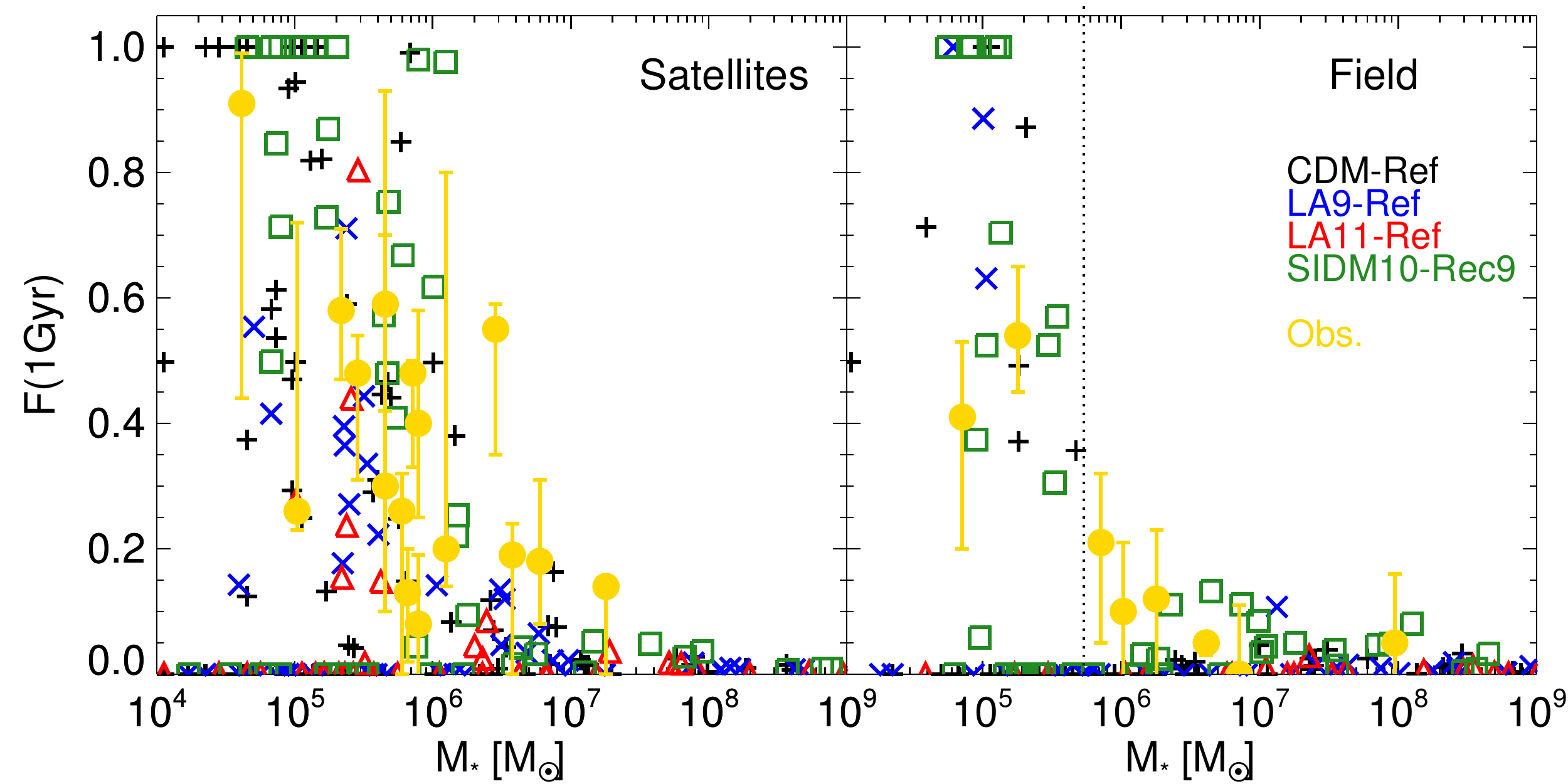}
     \caption{The stellar mass formed in the first gigayear in different dark matter models. We plot this fraction as a function of stellar mass, for MW and M31 satellites (left-hand panel) and for field galaxies within 2.5~Mpc of the M31-MW barycentre (right-hand panel). CDM-Ref galaxies are plotted as black plus signs, LA9-Ref as blue crosses, LA11-Ref as red triangles, and SIDM10-Rec9 as green squares. We plot observations from the compilation of \citet{Digby19} as gold circles, and in the field case we only include observational results that have been derived from the main sequence turnoff. These masses are measured at $z=0$, and therefore take account of stellar mass loss over the whole evolution of each galaxy.}
     \label{fig:G1vObs}
 \end{figure*}

 We stated above that the stellar mass formed within the first gigayear is accessible to observations \citep{Digby19}. In practice, the constraining power of this type of data set is limited due to both systematic uncertainties in the baryonic physics (c.f. differences between the APOSTLE and Auriga results presented in \citealp{Digby19}) and also the difficulty in measuring the ages of old stellar populations with an accuracy better than a Gyr. Nevertheless, it remains one of the best probes of star formation at very high redshift, and thus we consider it here as a first order, simple test that our models need to meet. We replicate Fig.~5 of \citet{Digby19} with our simulation results, that is, we plot the fraction of the $z=0$ stellar mass that was formed in the first gigayear as a function of stellar mass, together with the \citet{Digby19} compilation of observations, in Fig.~\ref{fig:G1vObs}.    
 
Both CDM and SIDM10 populate the regions that encompass the observational field and satellite data, with fractions at stellar masses $\ge2\times10^{6}\msun$ that are lower than 0.2 and fractions approaching 1 at $M_{*}<2\times10^{6}\msun$. The fractions for SIDM10 are higher than in CDM because although the stellar mass formed in the first gigayear is the same as in CDM (Fig.~\ref{fig:G1MMod}), the stellar mass formed at later times is suppressed relative to CDM (Fig.~\ref{fig:MsMod}). In line with the results from Fig.~\ref{fig:G1MMod}, the fraction of LA9 and LA11 stars formed in the first gigayear is lower than in CDM: the median fraction for LA9 satellites with total $M_{*}=[10^{5},10^{6}]\msun$ that do form any stars before 1~Gyr is 0.35, whereas for CDM that figure is 0.45. By contrast, none of the LA11 field galaxies exhibits a mass fraction higher than a few per~cent and is therefore at first glance in strong disagreement with the data. The LA9 simulation does feature three field galaxies with fractions higher than 0.5 -- but still far fewer than the 11 observed systems with fractions above 0.1. We therefore conclude that there are measurable differences between the models, but CDM, SIDM10 and LA9 trends are broadly consistent with current observations especially considering the difficulties in modelling both the simulations and the observational data; a more careful analysis that addresses these challenges has the possibility of ruling out LA11.

\subsubsection{The onset of star formation}
\label{sec:oosf}
 
Accurately determining the stellar mass formed at these early times depends in part on how the interplay between dark matter and galaxy formation physics delays the onset of star formation. We present two, complementary methods to get further insight into the nature of this delay below: the condensation time, and the formation time of the first star particle.

The earliest time at which star formation is expected to occur via atomic hydrogen cooling is when the progenitor halo masses first exceed the mass required for atomic cooling, since the halo mass determines the halo virial temperature. We refer to this time as the halo `condensation time'. The time at which each simulated galaxy forms its first star particle -- or the oldest star particle -- is much more sensitive to the stochastic assembly of haloes, the properties of the galaxy formation subgrid model, and also the cooling model; within the EAGLE implementation, the last of these was developed with only the $z<4$ galaxy population in mind, and therefore omits some physics fundamental to the properties of high redshift, pristine gas. However, the time of formation of the first star particle still has value in that it reflects information from the effects of reionization radiation and also considers all of the progenitor haloes that coalesce to form the LG dwarfs, not just the most massive progenitor halo. We will therefore focus on the condensation time as our primary diagnostic, and use the oldest star particle as a secondary diagnostic, particularly when considering the impact of the reionization redshift.

\begin{figure*}
     \centering
     \includegraphics[scale=0.34]{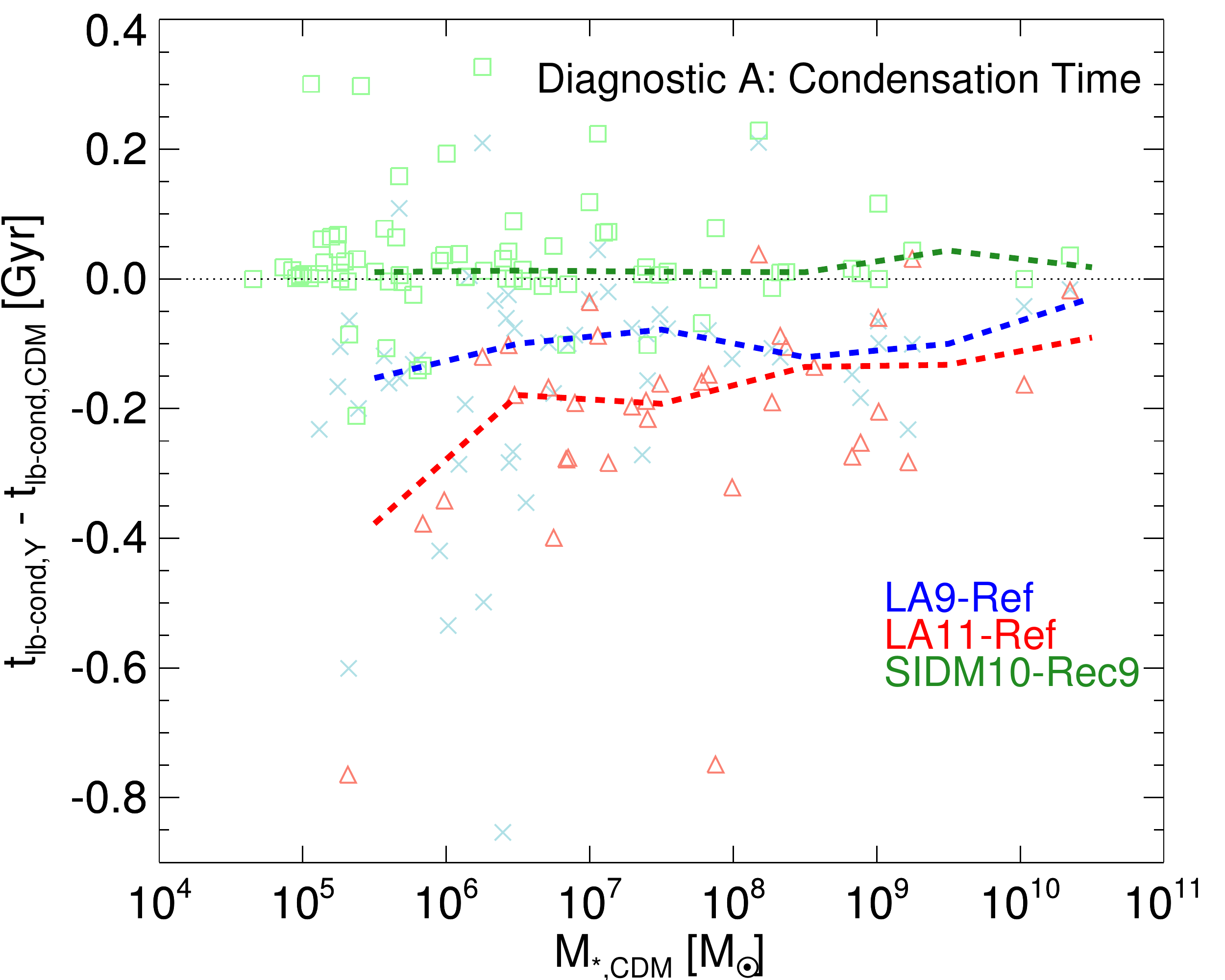}
     \includegraphics[scale=0.34]{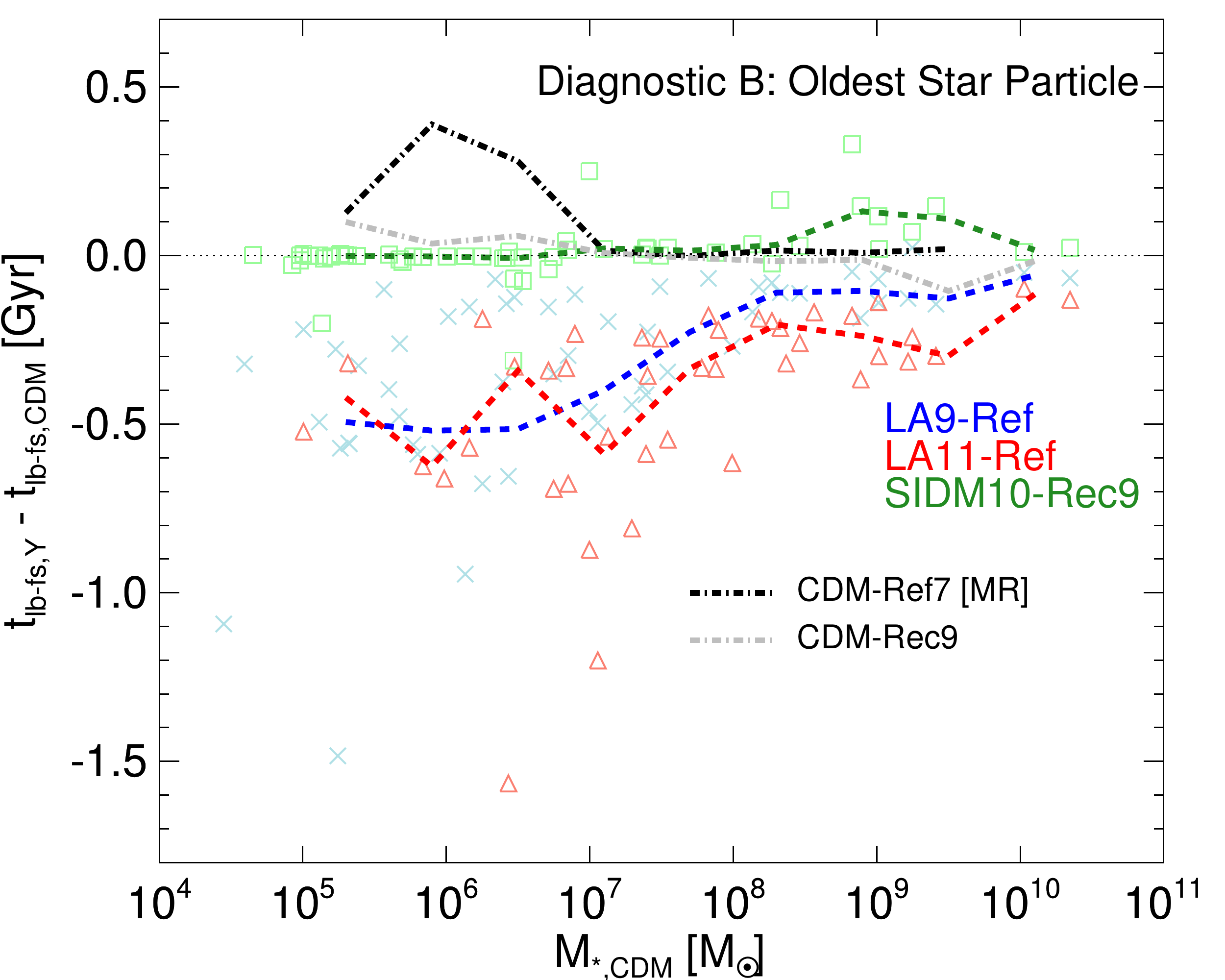}
     \caption{The difference in the condensation time (left-hand panel) and time of formation of the first star particle (right-hand panel). LA9 galaxies are shown as blue crosses, LA11 galaxies as red triangles and SIDM10 as green squares. Median relations are shown as dashed lines. The point colours are faded in order to make the median curves visible. The WDM galaxy $t_\rmn{lb}$ are matched to the CDM-Ref simulation, whereas the SIDM10-Rec9 $t_\rmn{lb}$ are matched the CDM-Rec9 $t_\rmn{lb}$; note that in this latter case the $x$-axis stellar mass is still that of the CDM-Ref counterparts to each CDM-Rec9 galaxy. In the right-hand panel we also show the median difference in first star formation time between the $z_\rmn{re}=7$ and fiducial ($z_\rmn{re}=11.5$) versions of CDM-MR as a black dot-dashed line and between CDM-HR-Rec9 and CDM-HR-Ref as dot-dashed grey line.}
     \label{fig:TvMMod}
 \end{figure*}

We consider the effect of the dark matter model on both of these diagnostics by computing the difference between the condensation times for CDM-alternative dark matter matched pairs, and show the results in Fig.~\ref{fig:TvMMod} as a function of AP-HR-CDM $z=0$ stellar mass (left-hand panel); we repeat the procedure in the same figure for the oldest star particle (right-hand panel). In the latter case, we also consider the role of reionization redshift with two additional runs: AP-MR-CDM-Ref7, which is matched to AP-MR-CDM (plotted as a function of the fiducial AP-MR-CDM $z=0$ stellar mass), and AP-HR-CDM-Rec9, which is matched to both AP-HR-CDM and AP-HR-SIDM10-Rec9 (in both cases plotted as a function of the counterpart fiducial AP-HR-CDM $z=0$ stellar mass).
 
 Both the condensation times and the oldest stellar populations are delayed in LA9 relative to CDM, and more so in LA11. There is evidence from both metrics that the delay is longer for less massive galaxies: the condensation time for LA9 (LA11) galaxies at $M_{*}\sim10^{8}\msun$ is delayed by of order 100~Myr (200~Myr) and at $10^{6}\msun$ by 300~Myr (400~Myr). The delay time is even longer for the oldest star particle: the median delay by this metric is 100~Myr (200~Myr) for LA9 (LA11) galaxies at $M_\rmn{*,CDM}=10^{9}\msun$ and 400~Myr (600~Myr) at $M_\rmn{*,CDM}=10^{7}\msun$. 
 
 The SIDM10 simulation instead shows very little difference from CDM by either metric, with the possible exception of an earlier formation time of the oldest star particle at $M_{*}>10^{8}\msun$. By contrast, the difference caused by the reionization redshift is dramatic for stellar masses below $10^{7}\msun$ -- the stellar mass of the brightest MW dwarf spheroidals -- with less massive galaxies forming their first stars progressively earlier when reionization takes place later than $z=11.5$ \citep[see also][]{GarrisonKimmel19}. Early reionization pumps energy into the IGM at earlier times, delaying cooling until more massive haloes can form; in evacuating gas from haloes their growth rates are slowed down. When coupled with the uncertainties in how cooling proceeds in pristine gas in the early Universe \citep{Munshi19}, it is clear that discerning the degree to which the initial onset of star formation is set by heating from reionizing background, versus the presence of a cut-off in the matter power spectrum, will require galaxy formation models tailored to the needs of high redshift science. 
 
   \subsubsection{Environment}
  
    We have shown so far that the onset of star formation is set in part by the location -- or absence -- of a cut-off in the matter power spectrum due to the particle nature of dark matter, by the mass of the galaxy, and by the redshift of reionization. A further dependence was revealed by \citet{Lovell19b} on environment; objects start forming later in underdense environments than in overdense ones, and this gradient in the ages of the oldest stars between overdense and underdense regions is stronger for dark matter models with a power spectrum cut-off. \citet{Lovell19b} showed that this was a $\sim$10~per~cent effect in the very different regime of a $z=6$ volume, and here we look for evidence of this phenomenon in our simulations, with a view to establishing whether it could be observed in the LG as a difference in age between satellites and field galaxies{, especially since no such difference is expected in the CDM model \citep{Dixon18}}.

\begin{figure}
     \centering
     \includegraphics[scale=0.33]{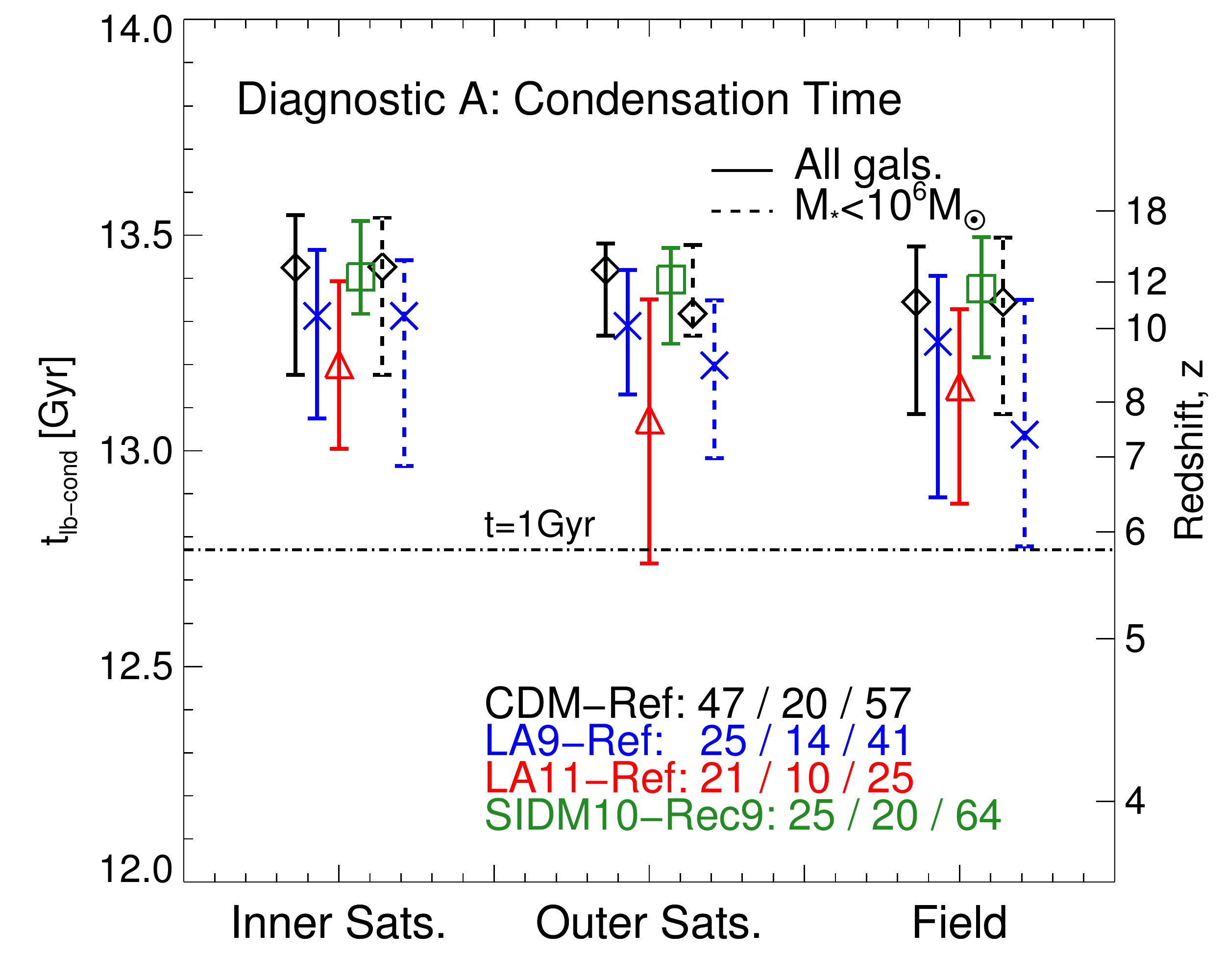}
     \includegraphics[scale=0.33]{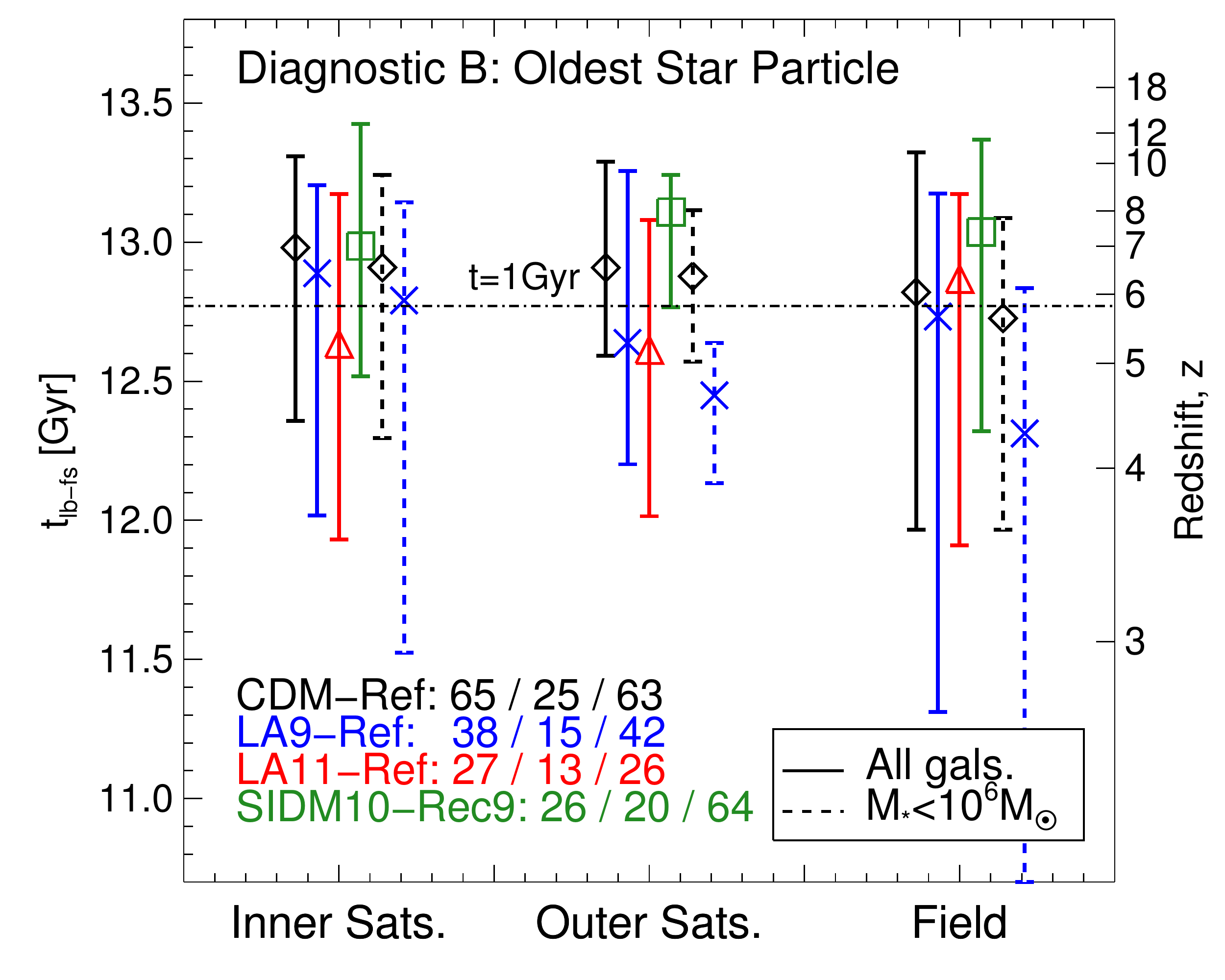}
     \caption{Environmental dependence of the delay in the onset of star formation. The top panel shows the distribution of times at which each halo first attains the halo mass required for atomic HI cooling, the bottom panel instead shows the distribution of the formation times of the first stellar population particle in each galaxy. We split the populations of galaxies into inner satellites, outer satellites, and field galaxies as described in the text and give the numbers in each category. The CDM results are shown as black diamonds, LA9 as blue crosses, LA11 as red triangles, and SIDM10 as green squares. Data points mark medians, and error bars the 68~per~cent region. The CDM and LA9 distributions are plotted for both the whole galaxy sample and for the subsample with $M_{*}<10^{6}\msun$, in the latter case using a dashed error bar. The dot-dashed line marks the first 1~Gyr, and the right-hand $y$-axis shows the redshift corresponding to each lookback time. Note the change in scale on the $y$-axes between the two panels.}
     \label{fig:SvBvF}
 \end{figure}

    We proceed as follows. We separate our simulated galaxies into three environment bins: inner satellites ($<150$~kpc from either of the M31 or MW analogue galaxies), outer satellites (150~kpc to 300~kpc from the M31 or MW analogues) and field galaxies (further than 300~kpc from either massive galaxy, and less than 3~Mpc from M31-MW barycentre); note that we do not take into account whether any of the `outer satellites' have passed through a pericentre with 150~kpc of the host centre or viceversa. For each of these environments we compute the median, the 16th percentile and 84th percentile for the condensation time and the age of oldest star particle diagnostics and present the results in Fig.~\ref{fig:SvBvF}. For all four models we show results computed when using all galaxies irrespective of stellar mass or of whether they have a match with a CDM halo; for the CDM and LA9 models we also include results for low mass galaxies ($M_{*}<10^{6}\msun$), in the expectation that the difference between the models may be more pronounced for faint dwarfs.       
    
 In all three environments, the condensation time shows a systematic shift from CDM to LA9 to LA11 by $\approx100$~Myr, as was previously shown for matched pairs in Fig.~\ref{fig:TvMMod}. This is similar to the delay found between ETHOS and CDM galaxies in \citet{Lovell19b}, and somewhat longer than the 10~Myr delay found for the progenitors of brightest cluster galaxies in WDM by \mbox{\citet{Esmerian19}}. The condensation time for CDM galaxies is consistent in all three environments, as is the case for SIDM10. However, there is some evidence that the LA9 and LA11 galaxies in the outer two bins experience a longer delay compared to CDM than do the inner satellite populations, particularly with the latest forming galaxies in each sample. This delay also appears to be a few hundred Myr longer in the faint galaxy population. For future comparisons to observations, like those in Fig.~\ref{fig:G1vObs}, one difference between LA9/LA11 and CDM is that some of the WDM galaxy populations touch the 1~Gyr line but the latter do not, so some WDM haloes will fail to produce any stars from HI cooling before the end of reionization independently of the details of baryonic physics, and in this EAGLE/APOSTLE model, are reliant on retaining some of their gas into the post-reionization era in order to become luminous. 
 
 In the lower panel of Fig.~\ref{fig:SvBvF} we repeat this exercise with the oldest star particle diagnostic. The scatter between galaxies is much larger due to stochasticity and non-linear baryon physics, and all of the models show a delay in star formation relative to the condensation time. Hints that this delay is sensitive to the reionization redshift are apparent when comparing CDM and SIDM10 (-Rec9): their haloes condense at the same time, but on average the first star particle -- at least for the outer satellite and field environments -- forms several hundred Myr earlier in the SIDM10-Rec9 run, with its lower redshift of reionization. All of the models rely on some of their haloes retaining gas beyond the end of reionization in order to form galaxies, especially for LA11 and for the faint LA9 galaxies. The WDM models still show a progressive delay in the formation of the first star particle with respect to CDM in the satellite populations, even though the majority of CDM and WDM haloes have condensed prior to the first star particle time. However, we stress once again that detailed models of the very high redshift interstellar and intergalactic media will be required to make precise predictions for this observable, together with radiative transfer calculations of the sort calculated by \citet{Dixon18} and \citet{Liu19} to follow the process of reionization in different environments.
 
 \subsubsection{Distribution of condensation times}
 
\begin{figure}
     \centering
     \includegraphics[scale=0.34]{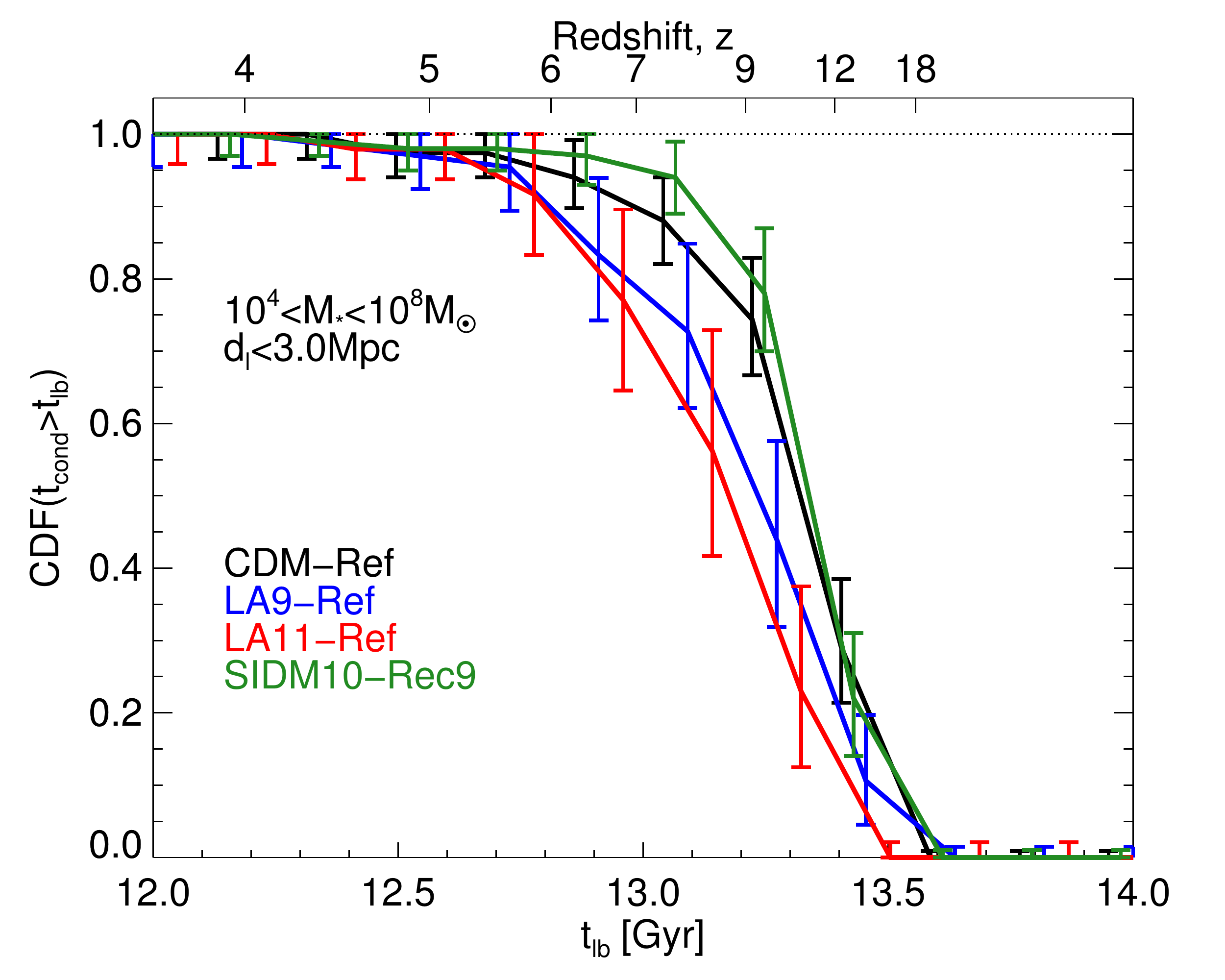}
     \caption{The fraction of galaxies whose condensation time is older than lookback time $t_\rmn{lb}$; the correspondence between model and colour is the same as in the previous figures. The error bars are binomial. We select galaxies with stellar masses in the range [$10^{4}$-$10^{8}\msun$]. }
     \label{fig:TofsCuml}
 \end{figure}

We end our presentation of the results with a prediction for the distribution of condensation times in each model. In Fig.~\ref{fig:TofsCuml} we plot the cumulative distribution of condensation times for the CDM, WDM and SIDM10 LG galaxies with stellar mass in the range $10^{4}-10^{8}\msun$.  We include all simulated galaxies within the stated stellar mass range and within 3~Mpc of the MW-M31 axis, regardless of whether they have a CDM counterpart.   
 
\begin{figure}
     \centering
     \includegraphics[scale=0.34]{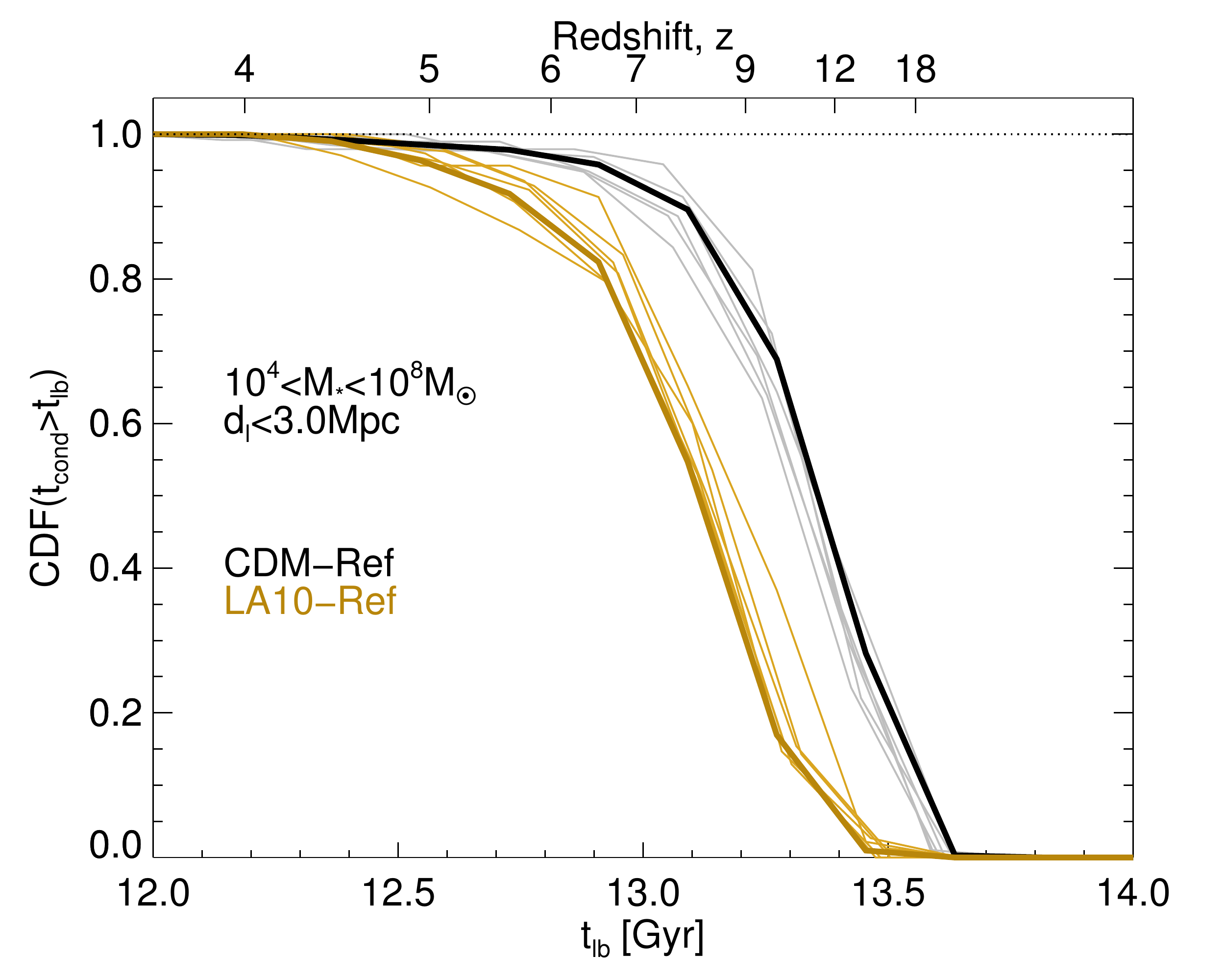}
     \caption{As Fig.~\ref{fig:TofsCuml}, but for CDM-MR (black curves) and LA10-MR (brown curves) in order to study the variance between volumes. Thin lines represent the six individual volumes, and thick lines the median across volumes. }
     \label{fig:TofsCumlV16}
 \end{figure}

The LA9 condensation time distribution is delayed relative to CDM, and as expected, for LA11 the delay is even greater. This difference is slight: the 50~per~cent value for the LA9 galaxy distribution is only $\approx$300~Myr later than for CDM, and for the LA11 galaxies it is only less than 400~Myr later. By contrast, the SIDM10 condensation time distribution is similar to that of CDM, and perhaps even biased slightly towards earlier times. We speculate that this may be due to the later reionization time as discussed above (c.f. \citealp{GarrisonKimmel19}). We repeat this exercise for CDM and the LA10 model of sterile neutrinos, (see Table~\ref{tab:sims}) with the goal of obtaining a first-order understanding of how much the condensation times distribution varies between simulation volumes. We present the cumulative distribution of the condensation times in Fig.~\ref{fig:TofsCumlV16}, both for  medium resolution simulations of individual volumes as thin lines and the medians of the two models across all volumes as thick lines.  
 
 CDM and LA10 show similar behaviour for $t_\rmn{lb}<12$~Gyr, and then separate at higher lookback times by up to 500~Myr. The spread in the condensation time at which 50~per~cent of the galaxies have attained the HI cooling mass for both models is somewhat smaller, $\lsim400$~Myr, and we note that there is no overlap between the LA10 distributions and CDM. However, we anticipate that, once these distributions are convolved with the stochasticity of star formation, a degeneracy will be introduced between the oldest stellar populations of later-forming CDM systems versus earlier-forming LA10 systems. Finally, we note that the median curve in Fig.~\ref{fig:TofsCumlV16} would lie almost on top of the high resolution CDM simulation from Fig.~\ref{fig:TofsCuml} so the uncertainty due to resolution is smaller than that due to the dark matter model or the scatter between volumes.
 
 \section{Summary \& conclusions}
 \label{conc}
 
Various astronomical observables may yield insights into the nature of the dark matter. One such signature that has received attention in recent years is the impact on the distribution of star formation histories of dwarf galaxies, and in particular the time at which star formation first begins, which can be examined simultaneously in galaxies detected during the epoch of reionization \citep[][]{hashimoto18} and very old stellar populations in LG galaxies \citep{Digby19}. It is well established that low mass haloes form later if there is a small scale suppression in the matter power spectrum at dwarf galaxy scales \citep{Lovell12,Ludlow16,Bose17a,WangL17}, a phenomenon that has been probed using estimates of the reionization redshift and optical depth \citep{Bose16c,Rudakovskyi16,Lovell18a}, plus the ages of local dwarf galaxies \citep{Lovell17b,Bozek19,Maccio19}. \citet{Lovell19b} moved beyond the generic star formation histories to consider the specific case of the oldest stellar populations as observed during the epoch of reionization; that work specifically considered the ETHOS framework of alternative (or interacting) dark matter models, which features both self-interactions and a cut-off in the matter power spectrum. In the present study we set out to examine the role of the power spectrum cutoff and self-interactions in a Local Group (LG) setting, on the star formation histories and on the populations of the oldest stars extant at $z=0$.

We considered different models of dark matter, and how they affect the properties of old stellar populations in the APOSTLE simulations of LG-analogue systems \citep{Fattahi16,Sawala16a}. We considered three models of resonantly produced sterile neutrinos: each has the same particle mass -- 7~keV -- but different lepton asymmetry, $L_{6}$ -- 9 (LA9), 10 (LA10) and 11.2 (LA11) -- all simulated with the same galaxy formation model and initial conditions, and also all consistent with the sterile neutrino decay interpretation of the otherwise unexplained 3.55~keV line \citep{Boyarsky14a,Bulbul14}. These models behave as warm dark matter (WDM). We also simulated a self-interating dark matter (SIDM) model with a velocity-independent self-interaction transfer cross-section per unit mass of 10~$\rmn{cm}^{2}\rmn{g}^{-1}$ (SIDM10), which was simulated with the same initial conditions phases as the CDM and WDM models but with slightly different galaxy formation model parameters, including a later redshift of reionization ($z_\rmn{re}=9$ rather than the fiducial $z_\rmn{re}=11.5$). We used the Lagrangian region matching algorithm of \citet{Lovell14} and \citet{Lovell18b} to identify counterpart haloes between different simulations of the same volume, both between alternative dark matter models and between different resolution simulations of the same model. We presented a first look at the matter distribution of the simulations, and showed that our alternative dark matter simulations preserve the large scale structure of the original CDM simulations (Fig.~\ref{fig:DM_Images}) and suppress the abundance of satellite galaxies, in the SIDM10 case likely through self-interactions between satellites and the host halo (Fig.~\ref{fig:Star_Images}). 

 We began our analysis with a study of how changing the resolution of the simulation affects the halo mass, total stellar mass, and the $z=0$ stellar mass that was formed in the first gigayear after the Big Bang. The halo masses typically varied by 10~per~cent for $M_\rmn{h}>10^{9}\msun$; however, the systematic variation was only 1~per~cent (Fig.~\ref{fig:MsRes}). The stellar masses were instead very strongly affected by resolution: relative to HR, the stellar mass is enhanced by up to a factor of 2 for the Milky Way (MW) and M31 analogue galaxies but suppressed by a factor of 2 for dwarf galaxies. No such deviations are apparent for the stellar mass formed in the first gigayear, although the statistical scatter is large (Fig.~\ref{fig:G1MRes}). We note that the variation with resolution were largely the same for CDM, LA11, and SIDM10 even though the latter features a different set of galaxy formation model parameters.

 We showed that the total stellar mass of galaxies with $M_{*}\geq10^{8}\msun$ is similar for WDM and CDM runs, but for lower mass galaxies both the total $z=0$ stellar mass and the $z=0$ stellar mass formed in the first gigayear after the Big Bang are suppressed in WDM relative to CDM, and in a manner that is stronger for the warmer of the two models; the LA11 stellar mass is suppressed to less than half of the counterpart CDM $M_{*}$ at $10^{6}\msun$ (Fig.~\ref{fig:MsMod}). We examined the star formation histories of a sample of galaxies that were matched between all four models (Figs.~\ref{fig:SFH_AM} and \ref{fig:SFH_AMdC}). The SIDM10 galaxies always formed their stars early, with later star formation suppressed relative to CDM. The WDM models showed a smaller degree of suppression than SIDM10, and we found some evidence of monolithic collapse starbursts. For the cumulative stellar mass formed in the first gigayear, the suppression in LA11 was still larger, by as much as a factor of 10 (Fig.~\ref{fig:G1MMod}). The SIDM10 first gigayear stellar masses showed better agreement with CDM despite the different choice of galaxy formation model. We showed that all four models can match the first gigayear actual stellar masses measured in LG satellites, but the LA11 model struggles to match the distribution of first gigayear masses in the field (Fig.~\ref{fig:G1vObs}).

The onset of star formation from atomic hydrogen cooling is of particular interest, as it will be limited by the time at which haloes collapse. We used two parametrizations for this time: first, the time at which each halo first passes the mass threshold for atomic cooling  -- the condensation time -- and second the age of the oldest star particle in each galaxy in order to show the effect of the redshift of the onset of reionization. These two diagnostics correlate with the position of the matter power spectrum cut-off, with LA9 galaxies forming their first stars later than their CDM counterparts and LA11 galaxies still later than LA9 (Fig.~\ref{fig:TvMMod}). The delay also correlated weakly with stellar mass: the median delay in condensation time relative to CDM at $10^{9}\msun$ was $\sim150$~Myr for both WDM models, increasing to 200~Myr for LA11 at $10^{7}\msun$. At the same time we showed that different reionization histories can change the time of onset of star formation to a similar degree. It will therefore be crucial that future studies model high redshift gas physics accurately in order to make precise predictions. By contrast, the SIDM10 galaxies showed very little difference in formation time to their CDM counterparts when simulated with the same galaxy formation model.  

We considered the distributions of condensation times in different environments and the scatter between systems. We showed evidence that field galaxies form marginally later than close-in satellites of M31 and MW analogue haloes; the evidence that the delay is longer for WDM models than in CDM is weak (Fig.~\ref{fig:SvBvF}). The time at which half of the WDM galaxies -- summed across all environments -- have condensed using the atmoic cooling diagnostic is delayed by $\sim$300~Myr relative to CDM, which is smaller than the time resolution for measuring the ages of old stellar populations (Fig.~\ref{fig:TofsCuml}). The SIDM10 model showed slightly earlier condensation times than CDM, which we suspect is a secondary imprint of the later reionization redshift used in the SIDM10(-Rec9) simulation. Finally, we showed that the scatter between condensation time distributions in 6 different volumes of the medium resolution CDM and LA10 simulations is only somewhat smaller than the difference between the two dark matter models (Fig.~\ref{fig:TofsCumlV16}). Once a full baryonic treatment of high-redshift cooling is implemented it will likely introduce further stochastic effects, and we therefore anticipate that it may be difficult to distinguish between later-forming CDM systems and earlier-forming WDM versions. The necessary observational power to measure age differences at better than $\sim100$~Myr presents a further barrier progress, and is unlikely to be determined without a successor to the {\it James Webb Space Telescope} (JWST) \citep{Weisz19}.  
 
 We conclude that the star formation histories of LG dwarf galaxies are interesting observables that can help us discern the physical nature of dark matter, particularly before the epoch of reionization. Warmer dark matter models show delays to the onset of star formation, but self-interactions appear to have no effect on this observable. However, this delay can be mimicked by the presence of an earlier reionization field. Therefore, improvements to the measurement of the ages of dwarf galaxy stars will have to go hand in hand with developments in modelling reionization and high redshift gas cooling reliably in order for this set of observables to contribute to be useful constraints on the nature of dark matter.  
 
\section*{Acknowledgements}

 MRL would like to thank Alis Deason for useful conversations. MRL is supported by a COFUND/Durham Junior Research Fellowship under EU grant 609412. MRL and JZ acknowledge support by a Grant of Excellence from the Icelandic Research Fund (grant number 173929). ADL is supported by the Australian Research Council through their future fellowship scheme (project number FT160100250). AR is supported by the European Research Council (ERC-StG-716532-PUNCA) and the STFC (ST/N001494/1). WAH acknowledge the support  via research project "VErTIGO" funded by National Science Center, Poland, under agreement no 2018/30/E/ST9/00698. AF is supported by a EU COFUND/Durham Junior Research Fellowship (grant agreement no. 609412), and a Science and Technology Facilities Council (STFC) [grant number, ST/P000541/1]. CSF acknowledges support from the European Research Council
(ERC) through Advanced Investigator grant DMIDAS (GA 786910). This work used the DiRAC@Durham facility managed by the Institute for
Computational Cosmology on behalf of the STFC DiRAC HPC Facility
(www.dirac.ac.uk). The equipment was funded by BEIS capital funding
via STFC capital grants ST/K00042X/1, ST/P002293/1, ST/R002371/1 and
ST/S002502/1, Durham University and STFC operations grant
ST/R000832/1. DiRAC is part of the National e-Infrastructure. This project has also benefited from numerical computations performed at 
the Interdisciplinary Centre for Mathematical and Computational Modelling (ICM) University of Warsaw 
under grants \#no GA67-15, GA67-16 and  G63-3.

\section*{Data availability statement}
 The simulation data for AP-HR-LA9, AP-HR-SIDM10-Rec9 (also MR), AP-HR-CDM-Rec9 and AP-MR-LA11 are published for the first time in this paper. Researchers wishing to gain access to these data should contact the lead author at lovell@hi.is.

\bibliographystyle{mnras}

\bsp
\label{lastpage}

\end{document}